\pdfoutput=1

\documentclass[
aps,
prd,
superscriptaddress,,
preprintnumbers,
nofootinbib,
showpacs,
showkeys,
]{revtex4-2}

\usepackage{fourier}
\usepackage{amsmath}
\usepackage{amssymb}
\usepackage{amsfonts}
\usepackage{mathrsfs}
\usepackage{latexsym}
\usepackage{graphicx}  
\usepackage{tikzsymbols}         
\usepackage{dcolumn}  
\usepackage{color}
\usepackage{xcolor}  
\usepackage{url}
\usepackage{bm}  
\usepackage{comment}              
\usepackage[english]{babel}
\usepackage[utf8]{inputenc}    
\usepackage{hyperref}  

\definecolor{skyblue}{RGB}{23,111,193}
\hypersetup{
    colorlinks = true,
    linkcolor = {black},
    urlcolor = {skyblue},
    citecolor = {skyblue},
    linkbordercolor = {white},
    urlbordercolor = {white},
    citebordercolor = {white},
}

\newcommand{\journal}[2]{\href{http://dx.doi.org/#1}{#2}}
\newcommand{\arxiv}[2]{\href{http://arxiv.org/abs/#1}{[arxiv:#1 #2]}}

\definecolor{lime}{HTML}{A6CE39}
\DeclareRobustCommand{\orcidicon}{
	\begin{tikzpicture}
        \draw[lime, fill=lime] (0,0) circle [radius=0.2] 
        node[white] {{\fontfamily{qag}\selectfont \tiny ID}};
        \draw[white, fill=white] (-0.0625,0.095) circle [radius=0.007];
	\end{tikzpicture}
	\hspace{-2mm}
}

\foreach \x in {A, ..., Z}
{\expandafter\xdef\csname orcid\x\endcsname{\noexpand\href{https://orcid.org/\csname orcidauthor\x\endcsname}
			{\noexpand\orcidicon}}
}

\begin{document}

\title{Minimal Leptogenesis in Brane-inspired Cosmology}

\author{Alessandro Di Marco\orcidA{}}
\email{alessandro.dimarco1@inaf.it}
\affiliation{National Institute for Astrophysics (INAF),\\
Institute for Space Astrophysics and Planetology (IAPS),\\
Via Fosso del Cavaliere, 100, 00133 Rome, Italy}

\author{Amit Dutta Banik\orcidB{}}
\email{amitdbanik@gmail.com}
\affiliation{Physics and Applied Mathematics Unit, Indian Statistical Institute, Kolkata-700108, India}

\author{Anish Ghoshal\orcidC{}}
\email{anish.ghoshal@fuw.edu.pl}
\affiliation{Institute of Theoretical Physics, Faculty of Physics, University of Warsaw, \\ 
ul. Pasteura 5, 02-093 Warsaw, Poland}

\author{Gianfranco Pradisi\orcidD}
\email{gianfranco.pradisi@roma2.infn.it}
\affiliation{University of Rome ``Tor Vergata'' \\
and INFN, Sezione di Roma ``Tor Vergata", via della Ricerca Scientifica 1, 00133 Roma, Italy}

\date{\today}

\begin{abstract}

We discuss how a post inflationary reheating phase characterized by a nonstandard multiple
scalar field cosmology can change the thermal history of the universe, affecting minimal high scale leptogenesis.  
In particular, we explore a class of models where a set of scalar fields in a brane-inspired dynamical scenario 
modifies the Boltzmann equations  concerning standard leptogenesis. 
The produced lepton asymmetry, due to the decays of heavy 
Majorana right-handed neutrinos responsible for generating Standard Model neutrino masses via the type-I seesaw, 
is affected as well. 

\end{abstract}

\maketitle
\tableofcontents

\section{Introduction}
\label{intro}

The theory of cosmological inflation \cite{1} represents an awesome solution
to the long-standing conundrums affecting the standard cosmology, i.e.
the flatness problem, the horizon problem and the monopole problem. Inflation also provides a natural explanation 
for the seeds, namely the primordial metric fluctuations generating the matter inhomogeneities responsible both for
the growth of the large-scale structures visible in the universe
and the temperature anisotropies of the Cosmic Microwave Background (CMB).  
In its simplest version, the so called single field slow-roll inflation, the inflationary mechanism is driven by
a homogeneous, neutral and minimally coupled scalar field $\phi$,
called the inflation field, typically characterized by an effective scalar potential $V(\phi)$ 
equipped with an (almost) flat region and a fundamental vacuum state.
In the first stage of inflation, the scalar field slowly crosses the plateau of the scalar potential, 
that behaves like a 
cosmological constant and triggers an (almost) de Sitter expansion of the universe.  
At the end of the inflationary period, 
the inflaton field reaches the steep region of the potential and 
falls in the global vacuum where it starts to oscillate. 
As a consequence, it should then decay to Standard Model (SM) and Beyond-the-Standard-Model (BSM) relativistic particles,
reheating the cold universe and giving rise to the graceful exit toward 
the standard initial radiation-dominated Hot Big Bang (HBB) epoch (for reviews on reheating see \cite{2}).  
Of course, the simplest mentioned scenario is not mandatory.
The universe could have experienced nonstandard post-reheating and 
pre-Big-Bang-Nuclesynthesis (pre-BBN) cosmological phases driven by one or more additional  
scalar fields, recovering the radiation dominance at lower energy scales. 
An intriguing possibility, first noticed in 
\cite{3}, is represented by the presence of additional sterile scalar fields characterized 
by a faster-than-radiation scaling law of the corresponding energy density. 
As new cosmological components, they can provide interesting modification of the 
dark matter annihilation rates and relics \cite{3,4,5,6,7}, 
inflationary e-folds \cite{8,9}, lepton and baryon asymmetry generation \cite{10,11}, 
matter-dark matter cogenesis \cite{12} and gravitational wave signals \cite{13}. 

These scalars are common in theories with extra dimensions and branes \cite{14}, 
like superstring orientifold models \cite{15}. 
Indeed, scalars parametrizing the positions of D-branes along the transverse internal directions interact gravitationally 
with the metric sector and can be engineered in a way to be decoupled by the longitudinal oscillation modes, 
related to SM and BSM fields.  In this paper, we consider non-standard cosmologies inspired 
by orientifolds with D-branes containing, generically, 
multiple sterile scalar fields entering a non-standard post-reheating phase. 
We study their effects on minimal thermal leptogenesis, extending the analysis of \cite{10}, 
where the single-field case has been addressed.

The term leptogenesis refers to the process of generation of lepton asymmetry and (induced) baryon asymmetry in the universe. 
The simplest class of models employs the decay of heavy right handed neutrinos (RHNs) in type-I seesaw mechanism \cite{16}. 
The process involves CP violating out of equilibrium decay of lepton number violating RHNs.  
There are also several other versions of leptogenesis, depending for instance on the choice of the used seesaw mechanism 
(type-II, type-III), in the presence of supersymmetry (soft leptogenesis) \cite{17} 
or even by radiative generation \cite{18}. 
Complete reviews on leptogenesis can be found in \cite{19}.

We limit ourselves to the analysis of the effects of the mentioned insertion of multiple scalar fields 
on the minimal type-I seesaw leptogenesis, 
although we expect that similar modifications could be applied to other leptogenesis scenarios as well.

The paper is organized as follows. 
In Section II we briefly discuss non-standard cosmology in the post-reheating early universe epoch,  
with the presence of scalar fields inspired by properties of  (super)string theory vacua. 
In Section III we discuss how standard thermal leptogenesis is modified by the presence of a bunch of $k$ scalar fields, 
discussing the two-scalar-field case in details. 
In Section IV, we repeat the study of non-standard leptogenesis in the presence of three active scalar fields. 
Finally in Section V, we summarize our obtained results and discuss some open problems.
In this paper we use the particle natural units $\hbar=c=1$, we indicate with $M_P = 1/\sqrt{8\pi G_N}$
the reduced Planck mass where $G_N$ is the gravitational Newton constant.

\section{D-Brane scalars and non-standard cosmology}
\label{Dbrane}

In the standard HBB scenario, the universe after reheating should experience a very hot and dense radiation dominated era. 
In that phase, the evolution of the universe is well described by a homogeneous fluid that obeys the 
Friedmann equation

\begin{equation}\label{raddom}
    H^2(T)\simeq\frac{1}{3M^2_{P}}\rho_{rad}(T), 
\end{equation}
where the radiation energy density at a $T$ temperature is

\begin{equation}\label{rhorad}
    \rho_{rad}(T)=\frac{\pi^2}{30}g_{E}(T) T^4
\end{equation}
where $H = \dot{a}/a$ denotes the Hubble rate in a Friedmann-Lemaitre-Robertson-Walker (FLRW) metric whose $a(t)$ 
is the standard cosmic scale factor, 
while $g_{E}$ is the effective number of relativistic degrees of freedom, turning out to be

\begin{equation}
    g_{E}(T)=\sum_{b} g_{b}\left( \frac{T_b}{T}\right)^4 + \frac{7}{8}\sum_{f} g_{f}\left(\frac{T_f}{T}\right)^4, 
\end{equation}
where $b$ and $f$ label bosonic and fermionic contributions, respectively, 
while $T_b$ and $T_f$ indicate the corresponding temperatures.
However, since the reheating phase is largely unknown, 
there is room for many scenarios involving a graceful exit from inflation and a 
corresponding modified path to the HBB era of a radiation dominated universe. 
A simple possibility consists in a cosmological stress-energy tensor that, after reheating, 
could be dominated by non-interacting scalar fields
equipped with a faster-than-radiation dilution law of the corresponding energy density.
These kind of components are quite common 
both in scalar modifications  of General Relativity and in theories with extra dimensions. 
Among them, (super)string theories are (the only) consistent proposals for an UV quantum completion of General Relativity. 
Moreover, many scalars are naturally present in their  spectra. 
In four dimensions, the scalars result from dimensional reduction of ten dimensional fields and
parametrize the deformations of the internal (compact) manifolds.  
In orientifold models, that are genuine string theory vacua,  
additional scalar fields are related to the presence of D-branes, 
defects where open-string ends slide. 
Indeed, the position of space-time filling branes along the internal directions 
are additional parameters that correspond to scalar fields in the effective low energy action. 
At tree level, the scalars are moduli, i.e. their potential vanishes and their interactions are purely gravitational.  
In order to stabilize most of them, a known procedure is to introduce flux compactifications, 
namely adding vevs to some of the internal (form) fields in the spectra. 
This way, one also get (partial spontaneous) breaking of supersymmetry and back-reaction 
on the space-time geometry resulting into a warping of the metric.  
The generated scalar potentials are typically steep or dominated by kinetic terms, 
making the corresponding fields possibly active after the reheating phase. 
In particular, we consider a set of scalar fields that only interact with the inflaton 
but are completely decoupled from the rest.  
They correspond to positions in the transverse internal directions of a bunch of well 
separated\footnote{We are assuming that the branes move slowly in the compact space while 
the potential felt by their position give rise to a quick dilution due to the exotic nature of the corresponding fluid.} 
D-branes whose dynamics is decoupled from the visible sector (i.e. the SM fields), 
from other possible hidden (dark) matter sector and even from the fields related 
to the longitudinal degrees of freedom on the brane itself. 
Their mutual interactions can also be neglected and the position fluctuations 
do not interfere among themselves. 
One example of this kind of scenario has been given in ref \cite{8, 14}, 
where the transverse position of a probe D-brane behaves exactly as requested, 
once the DBI and the Wess-Zumino terms describing its dynamics are specialized to a warped geometry.  
As said, the most important point is that these scalar fields always interact with the inflaton, 
that can thus decay to them and to the remaining (relativistic) components of the standard reheating fluid. 
The previous conditions are necessary in order to avoid relics moduli fields 
that overclose the universe or ruin the BBN.

Let us thus analyze a modification of the evolution of 
the early universe after the reheating phase,
realized through the presence of the mentioned set of scalar fields 
$\phi_i (i=1,...,k)$ \cite{9}.  
They are assumed to dominate at different time scales until radiation 
becomes the most relevant component, 
well before the BBN era in order to guarantee the predictions about the light element abundances.
Given the assumptions, the total energy density after the inflaton decay can be assumed to be 

\begin{equation}\label{eqn:totalenergydensity}
    \rho_{tot}(T)=\rho_{rad}(T) + \sum_{i=1}^k \rho_{\phi_i}(T).
\end{equation}

We introduce the scalar fields in such a way that, for $i>j$, $\rho_{\phi_{i}}$ 
hierarchically dominates at higher temperatures over $\rho_{\phi_{j}}$ when the temperature decreases.
All the scalar fields, supposed to be completely decoupled 
from each other and from matter and radiation fields, can be described as perfect fluids diluting faster than radiation.
In this respect, the dynamics is encoded in 

\begin{equation}
    \dot{\rho}_{\phi_i} + 3H\rho_{\phi_i}( 1 + w_i ) = 0 , 
\end{equation}
where $w_i=w_{\phi_i}$ is the Equation Of State (EoS) parameter of the field $\phi_i$.
Integrating this equation one finds 

\begin{equation}
    \rho_{\phi_i}(T)=\rho_{\phi_i}(T_i)\left(\frac{a(T_i)}{a(T)} \right)^{4+n_i}, 
\end{equation}
with $n_i=3w_i -1$. 
The indices\footnote{It should be noticed that the $n_i$'s are not necessarily integers, 
even though in this paper we use for them integer values.} 
$n_i$, namely the ``dilution" coefficients, are understood to satisfy the conditions

\begin{equation}
    n_i>0, \quad n_i<n_{i+1}.
\end{equation}
$T_i$ can be conveniently identified with the transition temperature  
at which the contribution of the energy density of $\phi_i$ becomes subdominant with respect to the one of $\phi_{i-1}$.
In other words, the scalar fields are such that  

\begin{eqnarray}
    \rho_{\phi_i}>\rho_{\phi_{i-1}} \mbox{ for } T>T_i ,\\
    \rho_{\phi_i}=\rho_{\phi_{i-1}} \mbox{ for } T=T_i ,\\
    \rho_{\phi_i}<\rho_{\phi_{i-1}} \mbox{ for } T<T_i .
\end{eqnarray} 

Using the conservation of the ``comoving" entropy density

\begin{equation}
    g_S(T)a^3(T)T^3=g_S(T_i)a^3(T_i)T^3_i ,
\end{equation}
being $g_S$, defined by

\begin{equation}
    g_{S}(T)=\sum_{b} g_{b}\left( \frac{T_b}{T}\right)^3 + \frac{7}{8}\sum_{f} g_{f}\left(\frac{T_f}{T}\right)^3,
\end{equation}
the effective number of relativistic degrees of freedom associated with entropy, 
the energy density of the various fields at a temperature $T$ 
can be expressed in terms of the transition temperatures $T_i$ as 

\begin{equation}\label{ratioofrhophi}
    \rho_{\phi_i}(T)=\rho_{\phi_i}(T_i)\left( \frac{g_S(T)}{g_S(T_i)}\right)^{\frac{4+n_i}{3}}\left(\frac{T}{T_i}\right)^{4+n_i}.
\end{equation}

For the first scalar field $\phi_1$, by definition, the transition temperature coincides with 
that at the beginning of the radiation-dominated era, $T_1=T_{r}$, so that $\rho_{\phi_1}(T_1)=\rho_{rad}(T_1)$.
The second scalar field $\phi_2$ is subdominant compared to $\phi_1$ below the temperature $T_2$.  
Using Eq.\eqref{ratioofrhophi}
and observing that $T_2$ is the transition temperature at which $\rho_{\phi_2}(T_2)=\rho_{\phi_{1}}(T_2)$, one gets

\begin{equation}\label{eqn: energy phi2}
    \rho_{\phi_2}(T)= \rho_{\phi_1}(T_1)\left(  \frac{T_2 \,  g_S^{1/3}(T_2)}{T_1 \, g_S^{1/3}(T_1)} \right)^{4+n_1}  \left(  \frac{T\, g_S^{1/3}(T)}{T_2 \, g_S^{1/3}(T_2)}  \right)^{4+n_2} .
\end{equation}

This equation tells us that the energy density of the scalar field $\phi_2$ 
depends on the ratio between the two scales $T_1$ and $T_2$,
where the $\phi_1$-dominance occurs. 
In the same way, we can derive the analogous expressions for the other scalar fields. 
The energy density carried by the $i$-th field $\phi_i$ can thus be written as

\begin{equation}\label{eqn: general result}
    \rho_{\phi_i}(T)=\rho_{rad}(T_{r})\prod_{j=1}^{i-1}\left(\frac{T_{j+1} \ g_S^{1/3}(T_{j+1})}{T_j \ g_S^{1/3}(T_j)}\right)^{4+n_j}
    \left(\frac{T \ g_S^{1/3}(T)}{T_i \ g_S^{1/3}(T_i)}\right)^{4+n_i} , \quad i\ge 2 ,
\end{equation}
and, inserted in Eq. \eqref{eqn:totalenergydensity}, 
it allows to calculate the total energy density dominating the expansion of the universe after the standard reheating phase, 
up to the beginning of the radiation-dominated epoch. 
In particular, using eq. \eqref{rhorad}, 
one has\footnote{It should be noticed that the amplification parameter ${\mathcal{J}}^2(T)$ is the $\eta(T)$ parameter of ref. \cite{9}.} 

\begin{equation}\label{eq:totalrho}
    \rho(T) = \rho_{rad}(T) \ {\cal{J}}^2(T) ,
\end{equation}
where the (positive) ``correction factor'' determining the non-standard evolution in the presence of $k$ additional scalar fields results

\begin{align}\label{eq:complmodfact}
    {\cal{J}}^2(T) &= 1 +  \left(\frac{T_{r} \ g_E^{1/4}(T_{r})}{T \ g_E^{1/4}(T)}\right)^{4} \ \left(\frac{T \ g_S^{1/3}(T)}{T_1 \ g_S^{1/3}(T_1)}\right)^{4+n_1}\nonumber\\
    &+\sum_{i=1}^{k} \left(\frac{T_{r} \ g_E^{1/4}(T_{r})}{T \ g_E^{1/4}(T)}\right)^{4} \ \left(\frac{T \ g_S^{1/3}(T)}{T_i \ g_S^{1/3}(T_i)}\right)^{4+n_i}\ \prod_{j=1}^{i-1}\left(\frac{T_{j+1} \ g_S^{1/3}(T_{j+1})}{T_j \ g_S^{1/3}(T_j)}\right)^{4+n_j} .
\end{align}

Since, by assumption, there is not a change in the number of degrees of freedom between 
the end of the reheating phase and the beginning of the HBB phase, 
all the ratios of $g_S$ and $g_E$ at different temperatures are of order $1$.  
Thus, for sufficiently high $T$ and $k$ additional scalar fields, it turns out that (defining $n_0=0$) 

\begin{equation}\label{eq:modfactor}
    {\cal{J}}^2(T) \simeq 1 + \sum_{i=1}^{k} \, \prod_{j=1}^{i}\left(\frac{T}{T_j}\right)^{n_j-n_{j-1}} .
\end{equation}

As expected, the larger the number of additional scalar fields is, the larger is the correction factor.  
Typically, in string-inspired models one cannot have $k\rightarrow \infty$ 
because the number of scalar fields is related to the number of branes 
and to the geometric deformations of the internal compactification manifold, 
both limited by the rank of the gauge group and the number of extra-dimensions, respectively\footnote{Typically, ``before’’ moduli stabilization, one has $\mathcal{O}(100)$ moduli from the compactification manifold and a net number of $\mathcal{O}(30)$ branes. Of course, the number of brane moduli can be made arbitrary by putting (unstable configuration of ) brane-antibrane pairs.}.
Moreover, it is important to underline a couple of fundamental aspects.
First, the properties of these scalars 
(i.e. dilution parameters and transition temperatures)
cannot be completely arbitrary.
In particular, it should be guaranteed that the energy density at 
the production scale (the reheating epoch)
should not be larger than some cutoff $M$, bounded by the inflationary scale $M_{inf}$.
As a consequence, a corresponding strong bound on the reheating temperature $T_{reh}$ is demanded \cite{9}.
For instance, in the case of a single non-standard post-reheating scalar
with a dilution parameter $n_1$ and a transition-to-radiation temperature $T_1=T_r$, 
the necessary condition is just $\rho_{\phi_1}(T_{reh})\leq M^4$
that leads to the bound

\begin{equation}\label{eqn: bound_1}
    T_{reh} \leq \alpha_1 M \left(\frac{T_1}{M} \right)^{\frac{n_1}{4+n_1}}, \quad \alpha_1=\left(\frac{30}{\pi^2 g_E}\right)^{\frac{1}{4+n_1}} .
\end{equation}

In the case of a pair of non-standard post-reheating scalars
with $\phi_2$ dominating at higher temperature $T>T_2$ on $\phi_1$, 
the necessary condition becomes $\rho_{\phi_2}(T_{reh})\leq M^4$ at $T=T_{reh}$. 
As a consequence, one gets

\begin{equation}\label{enq: bound_2}
    T_{reh} \leq \alpha_2 M \left( \frac{T_1^{n_1} T_2^{n_2-n_1}}{M^{n_2}} \right)^{\frac{1}{4+n_2}}, \quad \alpha_2=\left(30/\pi^2 g_E\right)^{1/4+n_2}
\end{equation}
where, by assumption, $n_2>n_1$. Of course, similar expressions can be easily found for more than two additional scalar fields.
The second point we would like to stress is that
the presence of these additional early cosmological phases
typically alter the inflationary number of e-folds  \cite{8,9} with 
an extra contribution $\Delta N(\phi_i,T_{reh})$ proportional 
to the (logarithm of) ${\cal{J}}(T_{reh})$, i.e 

\begin{equation}
    N_*\sim \xi_* - \frac{1-3w_{reh}}{3(1+w_{reh})}\ln\left(\frac{M_{inf}}{T_{reh}}\right) 
    + \ln\left(\frac{M_{inf}}{M_{Pl}}\right)  + \frac{2}{3(1+w_{reh})}\ln{\cal{J}}(T_{reh}) ,
\end{equation}
where $\xi_*\sim 64$ and $w_{reh}$
is the mean value of the EoS parameter of the reheating fluid.
Thereby, this extra factor depends on the
additional setup of scalar fields (namely number of scalars and dilution indices) and on the properties of the reheating scale.
However, reasonable assumptions provides an \textit{enhancement}
of the number of $e$-folds of the order of $5$-$15$, 
also allowing refined predictions for most of the inflationary models.

\section{Non-Standard History of Leptogenesis with two scalar fields}
\label{leptogen}

In this section we probe the effects on leptogenesis of the described fast expansion of the universe with multiple scalar fields.  
We consider the simple type I seesaw mechanism including heavy Majorana RHNs 
that generate a lightest neutrino and induce lepton number violation. 
Complex Yukawa interactions with leptons result in  CP violation when the RHN decay processes are  
considered with loop mediated interactions. 
Finally, out of equilibrium decay of RHNs 
(or of the lightest RHN $N_1$, the so-called $N_1$ leptogenesis that we use here) 
produces Baryon asymmetry in the universe. 
The Lagrangian involving the process is given (for three generations) by

\begin{equation}\label{L}
    {\cal L}_{RHN}=-\lambda_{ik}\bar{l_i}\tilde{\Phi} N_{k} - \frac{1}{2}M_k\bar{N^c}_kN_k + h.c.\,   \ \quad i,k=1, 2, 3,
\end{equation}
where a diagonal flavor basis is selected for the RHNs.
The Standard Model Higgs doublet is denoted by $\Phi$, the corresponding conjugate is $\tilde{\Phi}$ while $l$ indicates a SM lepton doublet. 
With the above BSM extension, one obtains an active neutrino mass matrix

\begin{equation}\label{numass} 
    M_{\nu}=-m_{D}^TM^{-1}m_{D}\,\, ,
\end{equation}
where $m_{D}$ denotes the Dirac mass matrix with entries of 
order ${\cal{O}}\sim v_{\Phi}\lambda$ ($v_{\Phi}$ is the vacuum expectation value of the Higgs doublet)
and $M$ is the diagonal RHN mass matrix.
As mentioned, the amount of CP asymmetry generated in the process of $N_1$ 
decay for a hierarchical RHN mass distribution $M_3, M_2 >> M_1$ is measured by  

\begin{align}\label{CPasy}
    \epsilon & = \frac{\sum_{\alpha}[\Gamma(N_1 \rightarrow l_{\alpha}+\Phi)-\Gamma(N_1 \rightarrow \bar{l}_{\alpha}+\Phi^{*})]}{\Gamma_1}  \nonumber\\  
             &=-\frac{3}{16\pi}\frac{1}{(\lambda^{\dagger}\lambda)_{11}}\sum_{k=2,3}{\rm Im}[(\lambda^{\dagger}\lambda)^2_{1j}]\frac{M_1}{M_k}\,\, , 
\end{align}
with $\Gamma_1=\frac{M_1}{8\pi}(\lambda^{\dagger}\lambda)_{11}$ 
being the total decay width of lightest RHN $N_1$. 
The asymmetry parameter $\epsilon$ can be used to provide a limit on the $N_1$ mass via the
Casas-Ibarra (CI) parametrization formalism \cite{20}. 
Indeed it turns out that

\begin{equation}\label{limit} 
    |\epsilon|\leq \frac{3}{16\pi v_{\Phi}^2}M_1m_{\nu}^{max}\,\, ,
\end{equation}
with $m_{\nu}^{max}$ being the largest light neutrino mass. 
As a consequence, a lower bound (the Davidson-Ibarra bound \cite{21}, $M_1 \gtrsim 10^9$ GeV) 
emerges for the $M_1$ mass of the lightest RHN,  when neutrino oscillation parameters are taken into account.  
$N_1$ leptogenesis, effective at temperatures $T\gtrsim 10^{12}$ GeV, 
induces also a constraint on the reheating temperature after inflation at values  $T_{reh}>10^{12} $ GeV.
Disregarding the possibility of flavored leptogenesis, we limit ourselves to the usual thermal leptogenesis,  
solving for the simplified Boltzmann equations (BEs). 
In standard cosmology with a radiation dominated universe after reheating, they can be written as

\begin{align}
    &\frac{d Y_{N_1}}{dz} = -z\frac{\Gamma_1}{H_1}\frac{\mathcal{K}_1(z)}{\mathcal{K}_2(z)}\left(Y_{N_1}-Y_{N_1}^{EQ}\right)\,,\label{RHNBE}\\
    &\frac{d Y_{L}}{dz} = -\frac{\Gamma_1}{H_1} \left(\epsilon  z \frac{\mathcal{K}_1(z)}{\mathcal{K}_2(z)}(Y_{N_1}^{EQ}- Y_{N_1}) +  \frac{z^{3} \mathcal{K}_1(z)}{4} Y_L  \right)\, \, .\label{asy}
\end{align}

In Eqs.~\eqref{RHNBE}-\eqref{asy}, $Y_i=n_i/s$ 
denotes the abundance of the particle $i$, 
namely the ratio of its number density to the entropy density, 
while $Y_L=(Y_l-Y_{\bar{l}})$ is the lepton asymmetry.
The equilibrium abundance of the lightest RHN is \cite{19,22} 

\begin{equation}
    Y_{N_1}^{EQ}=\frac{45g}{4\pi^4}\frac{z^2\mathcal{K}_2(z)}{g_{S}}.
\end{equation}

It should be noticed that Eqs.~\eqref{RHNBE}-\eqref{asy} 
both depend on the modified Bessel functions ($\mathcal{K}_{1,2}$), 
on the Hubble parameter $H_1=H(T=M_1)=H(T) z^2$ and on the decay width $\Gamma_1$ of $N_1$ 
(or on the washout parameter $K=\frac{\Gamma_1}{H_1}$), 
while the BE for the lepton asymmetry also depends on the asymmetry parameter $\epsilon$. 
Solutions of these equations can be found in \cite{19}.
In the presence of multiple scalar fields as described in Section II, the above BEs for leptogenesis have to be modified.  
The case with a single additional scalar field can be found in \cite{10}. 
For simplicity, we consider explicitly the simplest two-scalar-field scenario. 
Modifications in BEs arise by the correction to the Hubble parameter (as derived in Sect.~\ref{Dbrane}). 
With the assumption of $g_S\sim g_E$ for large $T$, the total radiation density 
can be extracted from eqs. \eqref{eq:totalrho}-\eqref{eq:modfactor}, 
and results

\begin{equation}\label{rho_new} 
    \rho_{tot}(T) = \rho_{rad}(T)+\sum_i^2\rho_i(T)\, \\
    =\rho_{rad}(T)\left\{1+\left(\frac{T}{T_r}\right)^{n_1}\left[1+\left(\frac{T}{T_2}\right)^{(n_2-n_1)}\right]\right\},
\end{equation}
where $T\ge T_2$ corresponds to the epoch of $\phi_2$ scalar domination, $T_r\leq T\leq T_2$ represents that of  
$\phi_1$ dominated expansion while for $T\le T_r$ ($T_r=T_1$) the universe is fully dominated by radiation.
The modified Hubble parameter is thus

\begin{equation}\label{Hnew}   
    H_{new}=H\left\{1+\left(\frac{T}{T_r}\right)^{n_1}\left[1+\left(\frac{T}{T_2}\right)^{(n_2-n_1)}\right]\right\}^{1/2}\,
\end{equation}
and it gives rise to the following modified BEs

\begin{align}
    &\frac{d Y_{N_1}}{dz}= -z \, \frac{\Gamma_1}{H_1} \, \frac{1}{\cal{J}} \, \frac{\mathcal{K}_1(z)}{\mathcal{K}_2(z)}\left(Y_{N_1}-Y_{N_1}^{EQ}\right)\,, \label{RHNBE2}\\
    &\frac{d Y_{L}}{dz} =  - \frac{\Gamma_1}{H_1} \, \frac{1}{\cal{J}} \,  \left(\epsilon  z \frac{\mathcal{K}_1(z)}{\mathcal{K}_2(z)}(Y_{N_1}^{EQ}- Y_{N_1}) + \frac{z^{3} \mathcal{K}_1(z)}{4} Y_L  \right)\, .\label{asy2}
\end{align}

A convenient and useful way to write $\cal{J}$ is 

\begin{equation}\label{factor}
    {\cal{J}}=\left\{1+\left(\frac{M_1}{T_r z}\right)^{n_1}\left[1+\left(\frac{M_1}{T_r x  z}\right)^{(n_2-n_1)}\right]\right\}^{1/2}\, ,
\end{equation}
with $x=\frac{T_2}{T_r}$.
Looking into the modified BEs \eqref{RHNBE2} and \eqref{asy2}, 
it can be observed that, apart from the standard parameters $\epsilon$ and $K=\frac{\Gamma_1}{H_1}$, leptogenesis with
two scalar field depends on a set of four new parameters $n_1,~n_2~({\rm{or}}~n_2-n_1),~T_r/M_1$ and $x$ (or $T_2$),  
that naturally modify the abundance of lepton asymmetry $Y_L$ as compared to the one of standard leptogenesis. 
We solve numerically Eqs.\eqref{RHNBE2} and \eqref{asy2},  
considering two possible sets of initial conditions. 
The first, A, corresponds to the case where the abundance of RHN $N_1$ is the one at the equilibrium 
$Y_{N_1}^{in}=Y_{N_1}^{eq}$. 
The second, B, $Y_{N_1}^{in}=0$ is relative to the case in which the initial abundance of RHN vanishes.  
In both cases, we assume that lepton asymmetry is absent before the decay of $N_1$,  $~Y_L^{in}=0$.
In the next two subsections we discuss in details the solutions for 
the quantities involved in the modified BEs. 
Lepton asymmetry is partially converted into baryon asymmetry by sphalerons \cite{19}

\begin{equation}\label{transfer}
    Y_B=\frac{8n_f+4n_{H}}{22n_f+13n_{H}}Y_L\,\, .
\end{equation}

Note that $Y_B=\frac{28}{79}~Y_L$ (for $n_H=1, n_f=3$), 
consistent with the observed baryon asymmetry in the universe $Y_{B}=(8.24-9.38)\times 10^{-11}$ \cite{23}.

\subsection{Case $Y_{N_1}^{in}=Y_{N_1}^{EQ}$}

\begin{figure}
    \begin{center}
        \includegraphics[width=0.45\textwidth]{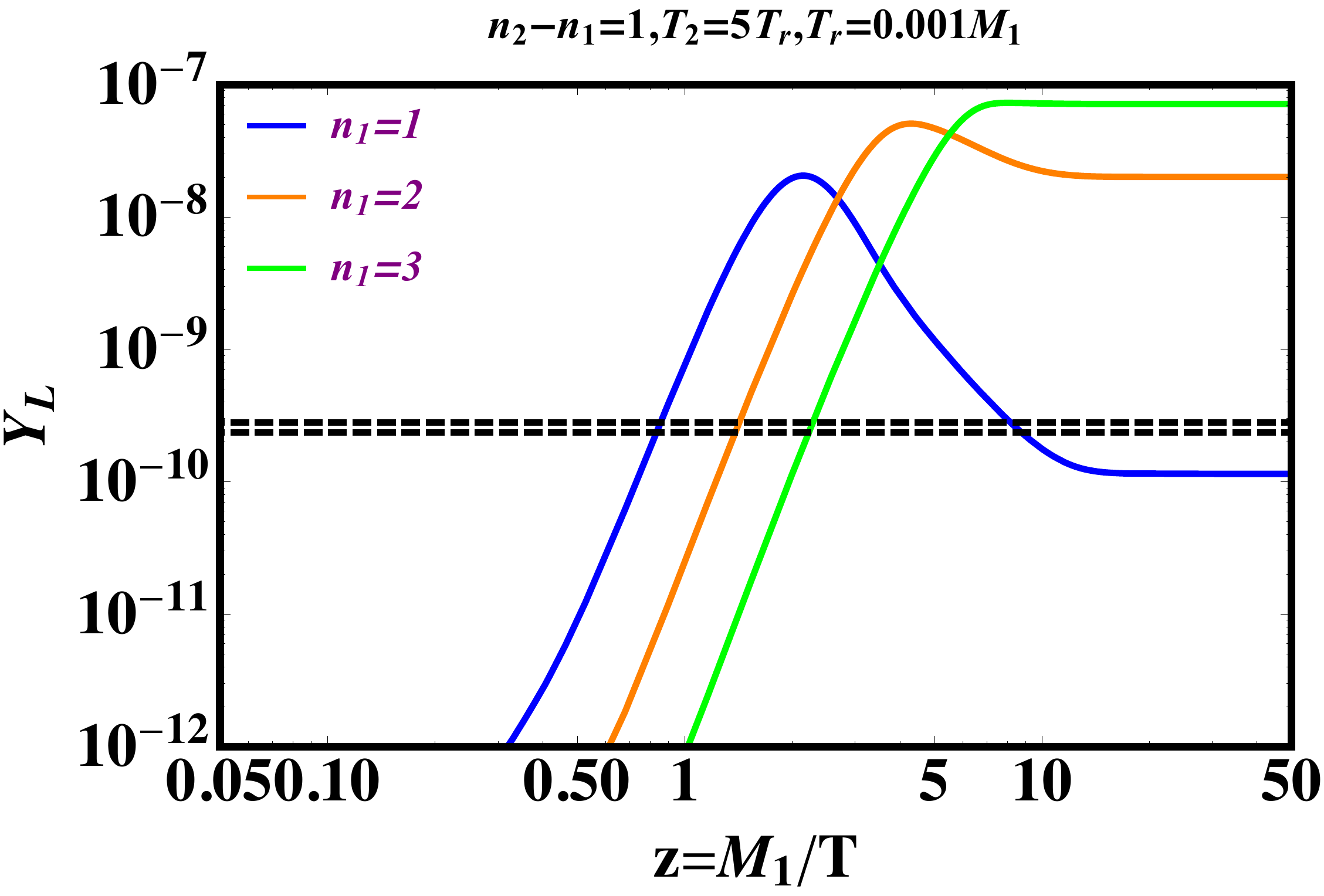}
        \includegraphics[width=0.45\textwidth]{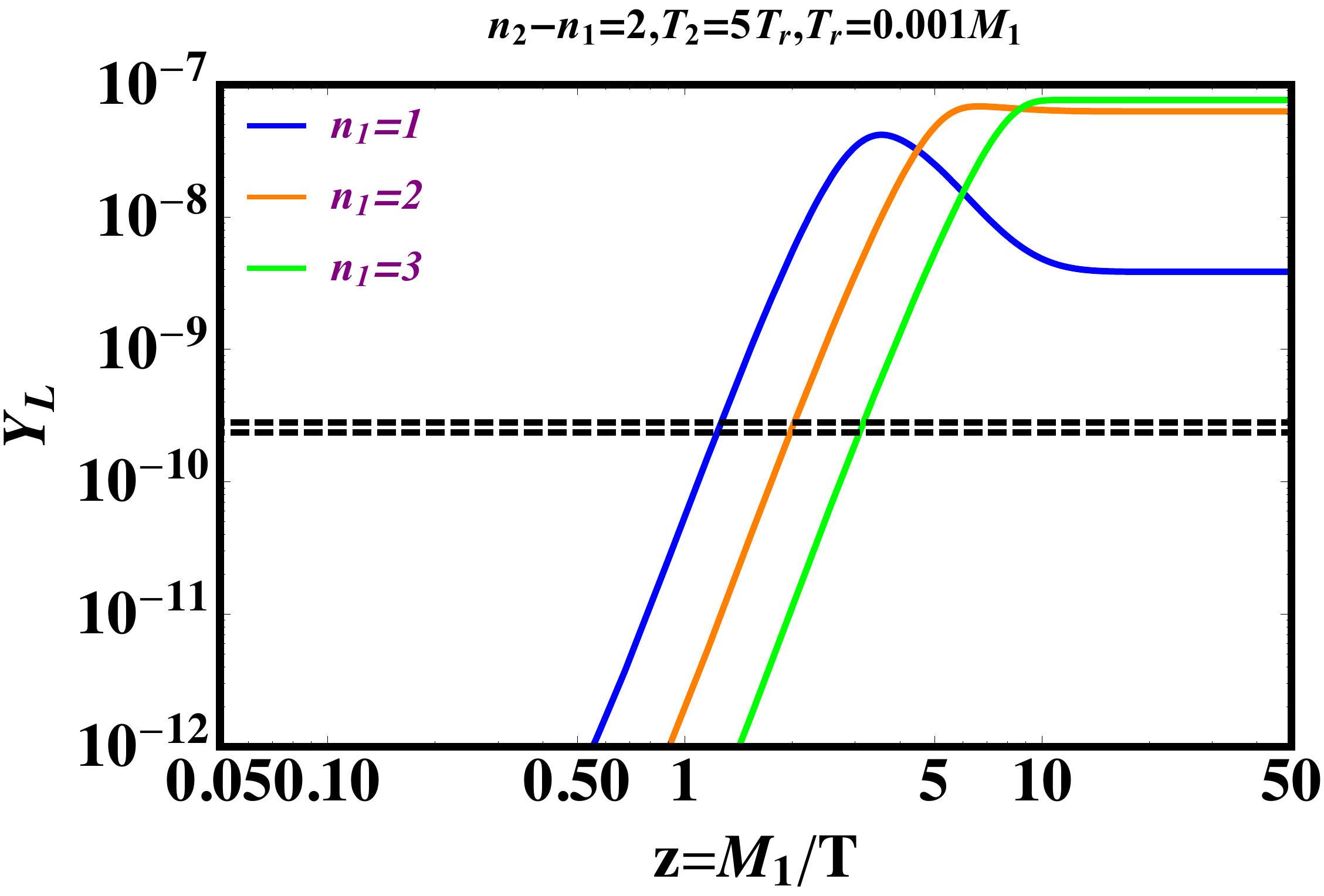}
        \caption{\it Evolution of $Y_L$ versus $z$ for initial equilibrium RHN abundance, $T_2=5 \, T_r$ and different $n_1$ values, with $n_2-n_1=1$ (left panel) and $n_2-n_1=2$ (right panel). The double black line(s) describe the baryogenesis threshold.}
        \label{fig:1}
    \end{center}
\end{figure}

In Fig.~\ref{fig:1} (left panel), we show how the lepton asymmetry 
$Y_L$ evolves as the universe expands with $z$, considering the modified Hubble parameter  
of eq.~\eqref{Hnew} for different $n_1$ but fixed $n_2-n_1=1$ values. 
For the purpose of illustration, other relevant parameters are kept fixed. 
In particular, $\epsilon=10^{-5}$, $(M_1=10^{11}~{\rm{GeV}})$, 
$K=\frac{\Gamma_1}{H_1}=600$, $T_r=10^{-3}M_1$
and $T_2=5T_r$. It can be observed a relevant increasing  
of $Y_L$ with the increasing of $n_1$, as expected because  
an higher $n_1$ corresponds to a faster expansion. 
Moreover, the increasing of $n_1$ also dilutes in a considerable way the washout effect, 
as manifested by the lowering of the inverse decay to $N_1$.

\begin{figure}
    \begin{center}
        \includegraphics[width=0.49\textwidth]{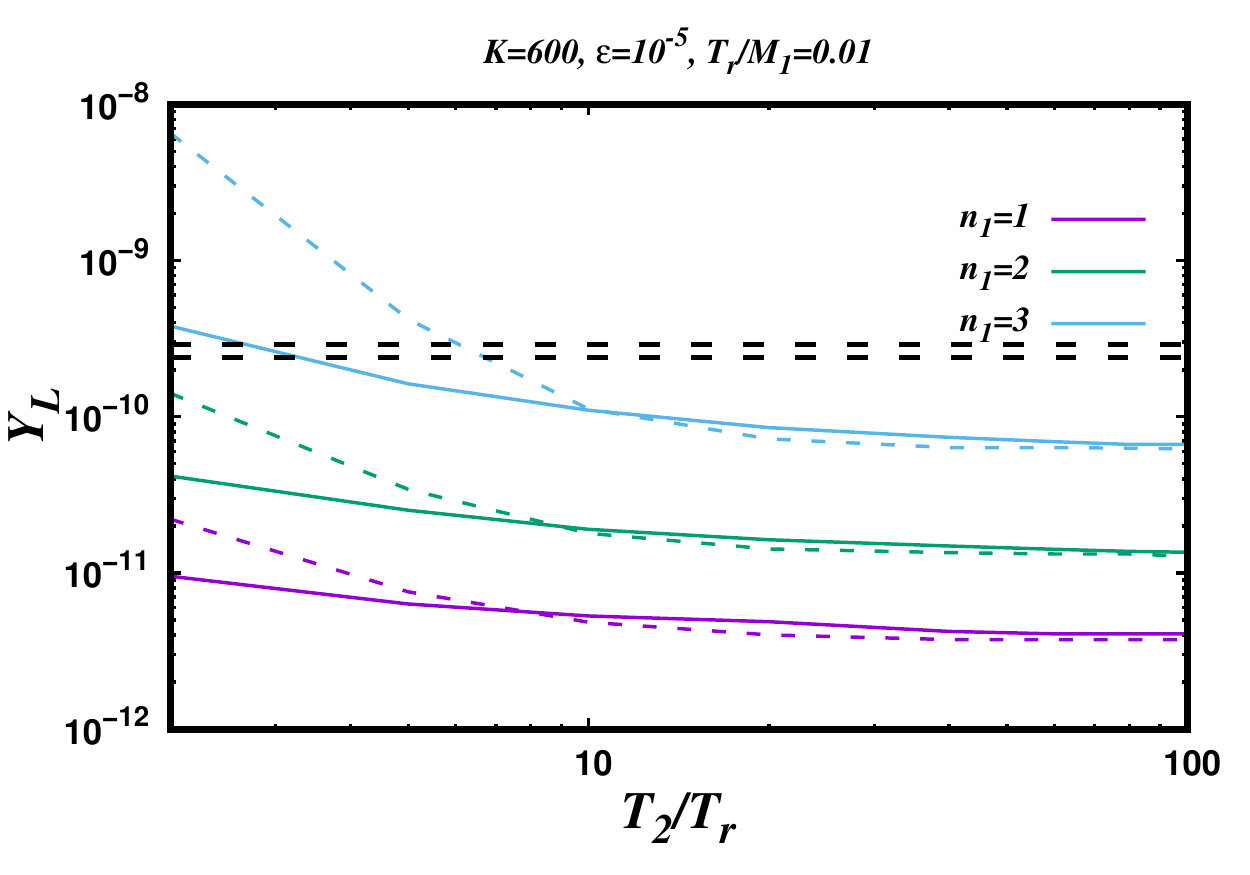}
        \includegraphics[width=0.49\textwidth]{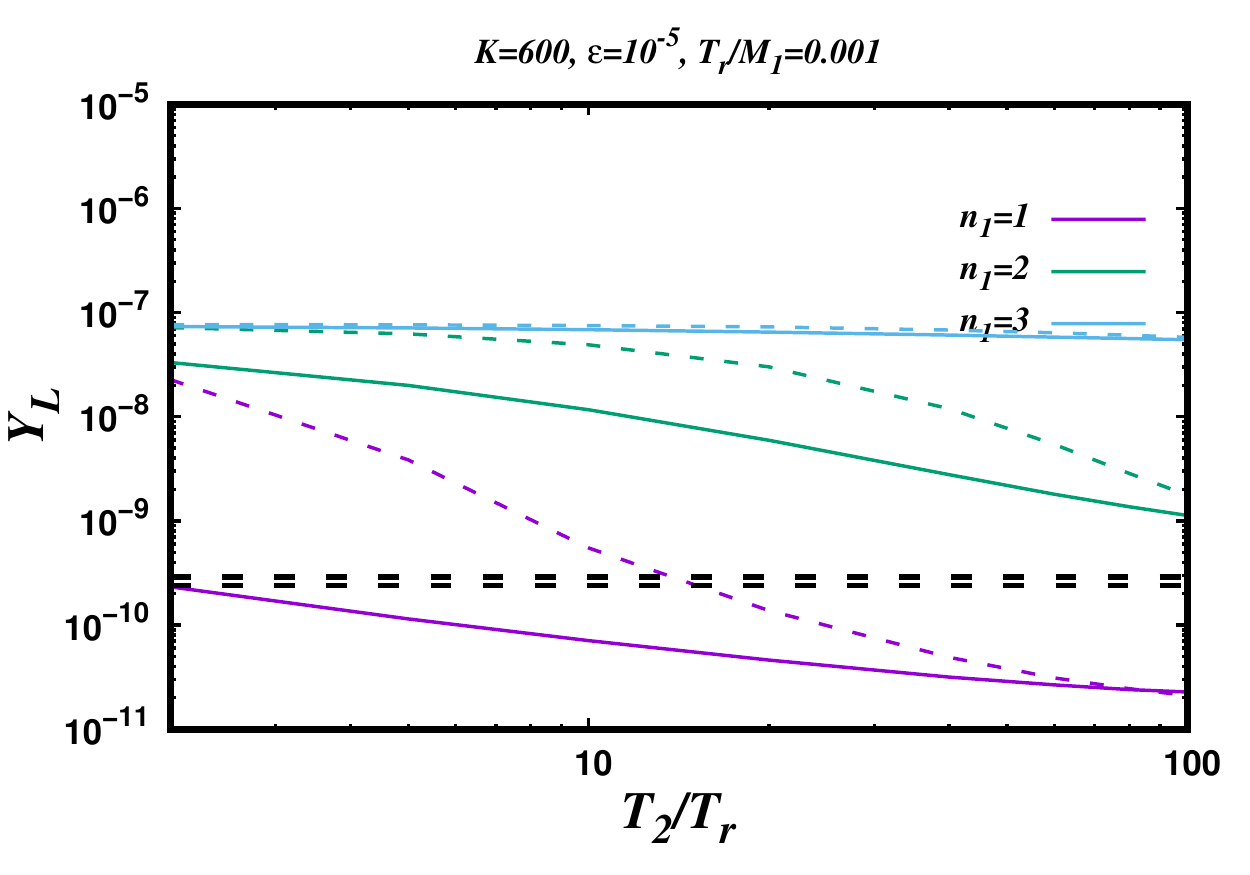}
        \caption{\it $Y_L$ versus $T_2/T_r$ for $T_r/M_1=0.01$ (left panel) and $T_r/M_1=0.001$ (right panel), with different $n_1$. Solid (dashed) lines refer to $n_2-n_1=1$ ($n_2-n_1=2$). The double black line(s) describe the baryogenesis threshold.}
        \label{fig:2}
    \end{center}
\end{figure}

In order to study the net effect of the presence of the second scalar, 
we consider a different $n_2-n_1=2$ in the right panel of Fig.~\ref{fig:1}, 
keeping the same set of values for the remaining parameters as in the left panel.  
Comparing the two plots, it emerges quite neatly that the second scalar field gives rise 
to an enhancement of the asymmetry $Y_L$ accompanied by a clear lowering of the washout effect. 
For high values of $n_1$, however, the second scalar is less important because 
the influence of $\phi_1$ is already very efficient, 
as demonstrated by the $n_1=3$ case where the fast expansion and the increasing of $Y_L$ 
lead to a negligible washout of the asymmetry.  
It is however worth to analyze the dependence on the other involved parameters.
For instance, it is very important the ratio between the temperatures separating the successive epochs of scalar domination.
To this aim, it is useful to plot the dependence of lepton asymmetry on the ratio of the two relevant temperatures $T_2$ and $T_r=T_1$. 
Results are reported in Fig.~\ref{fig:2} in the range $2\leq T_2/T_r\leq 100$, 
for the two different values of $T_r=10^{-2}~M_1$ and $T_r=10^{-3}~M_1$, 
keeping the  $\epsilon$ and $K$ values as in Fig.~\ref{fig:1}. 
In the left panel, the curves are referred to different values of $n_1$ and  $T_r=10^{-2}~M_1$, 
with solid lines related to $n_2-n_1=1$ and dashed lines to $n_2-n_1=2$. The same plot with $T_r=10^{-3}~M_1$ is reported in the right panel.
It happens that in the left-panel case the second scalar influences $Y_L$ only if $T_2\leq 10 \, T_r$, 
increasingly with increasing difference $n_2-n_1$, independently of the values of $n_1$. 
Notice that only for $n_1=3$ it is possible to get by leptogenesis the required baryon abundance in the universe (black bar).  
The behavior changes drastically when $T_r=10^{-3}~M_1$, as shown in the right panel of Fig.~\ref{fig:2}. 
Indeed, a decreasing in the value of $T_r/M_1$ reflects itself into an expansion of the epoch where $\phi_1$ 
dominates and a delay of the radiation domination. 
Thus, it helps to deviate the $N_1$ abundance from equilibrium and to generate lepton asymmetry.  
It can be easily noticed that the enhancement of $Y_L$ indeed allows to satisfy baryon asymmetry already for $n_1=1$, 
and there is higher sensitivity to the difference $n_2-n_1$, especially for large values of $T_2/T_r$, 
apart for the case $n_1=3$ when, as in the previous analysis reported in  Fig.~\ref{fig:1}, the washout is practically absent.
Finally, it is worth to observe that with the increasing of $T_2/T_r$, 
the effects related to the presence of $\phi_2$ become less and less important when the temperature decreases, 
becoming less prominent at the time of leptogeneis, as follows directly from eq. \eqref{factor}.  
For example, for $T_r/M_1=0.01$, the possible choice  
$T_2/T_r=100$ indicates that the influence of $\phi_2$ on the Hubble parameter ceases to exist at $T=M_1$ 
while, for  $T_2/T_r=10$, $\phi_2$ remains active up to $T=0.1 M_1$ altering the abundance of $Y_L$. 
As clear from \eqref{factor}, the second-scalar effect dominates for $M_1/T_r>>xz$, becoming insignificant if $xz\ge 100$.

\subsection{Case $Y_{N_1}^{in}=0$}

In this paragraph, we repeat the study of leptogenesis influenced by the presence of two scalars 
for vanishing RHN initial abundance (conditions ``B'').

\begin{figure}
    \begin{center}
        \includegraphics[width=0.45\textwidth]{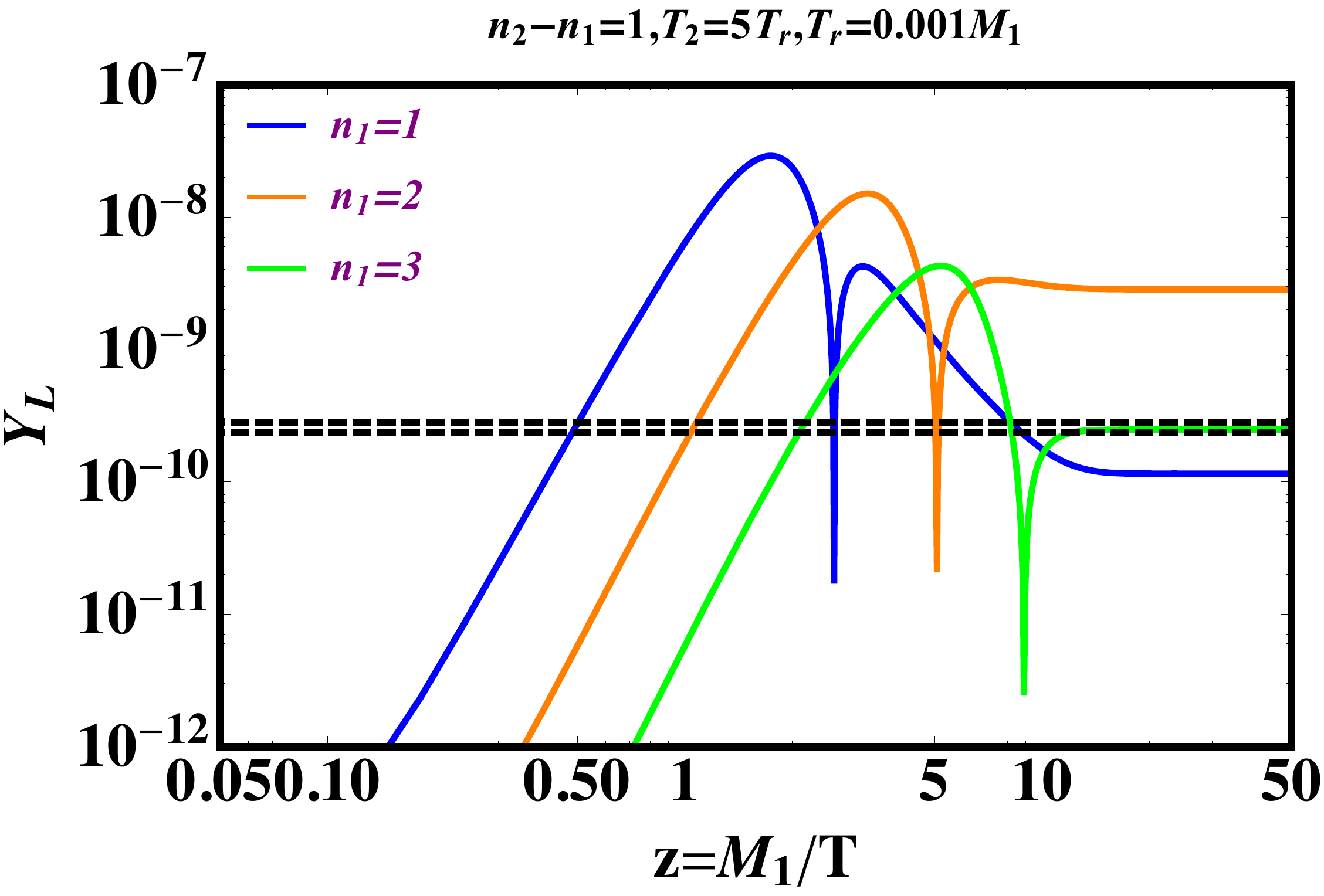}
        \includegraphics[width=0.45\textwidth]{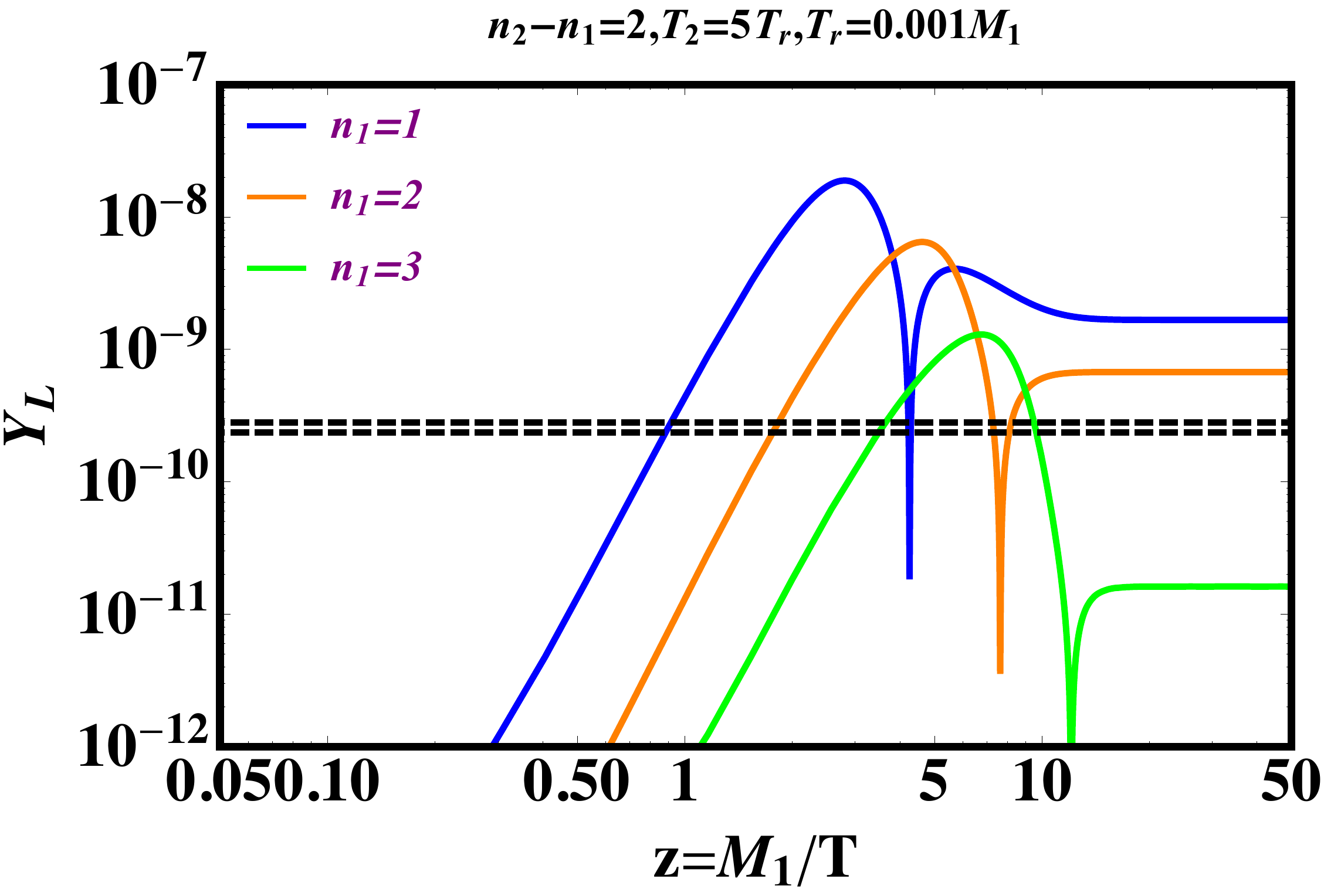}
        \caption{\it Same as Fig.~\ref{fig:1} for
        zero RHN abundance.}
        \label{fig:3}
    \end{center}
\end{figure}

In Fig.~\ref{fig:3}, we report the plots analogous to those in Fig.~\ref{fig:1} 
using the same set of parameters $\epsilon$, $M_1$, $T_r$, $T_2$, $K$ and $n_i$.  
In this scenario, the initially produced $N_1$ is 
then partially compensated by the inverse decay, 
resulting in an oscillation of negative lepton asymmetry giving later rise to the generation of a net positive lepton asymmetry.
It should be noticed that for flavourless leptogenesis
the Boltzmann equations give rise to solutions with a single
bounce on $Y_L$, while this is not the case
in more general frameworks where additional bounces
can occur, as shown in \cite{24}.
From Fig.~\ref{fig:3} (left panel) it turns out that an increasing value of $n_1$ from $1$ to $3$ 
provides a reduced washout of asymmetry resulting into an enhancement of the $Y_L$ value. 
However, for $n_1=3$, the $Y_L$ abundance value decreases significantly. 
This is due to the fact that a faster expansion also reduces the production of RHN by inverse decay. 
Similar effects have already been observed in the presence of a single additional scalar field \cite{10}.

\begin{figure}
    \begin{center}
        \includegraphics[width=0.49\textwidth]{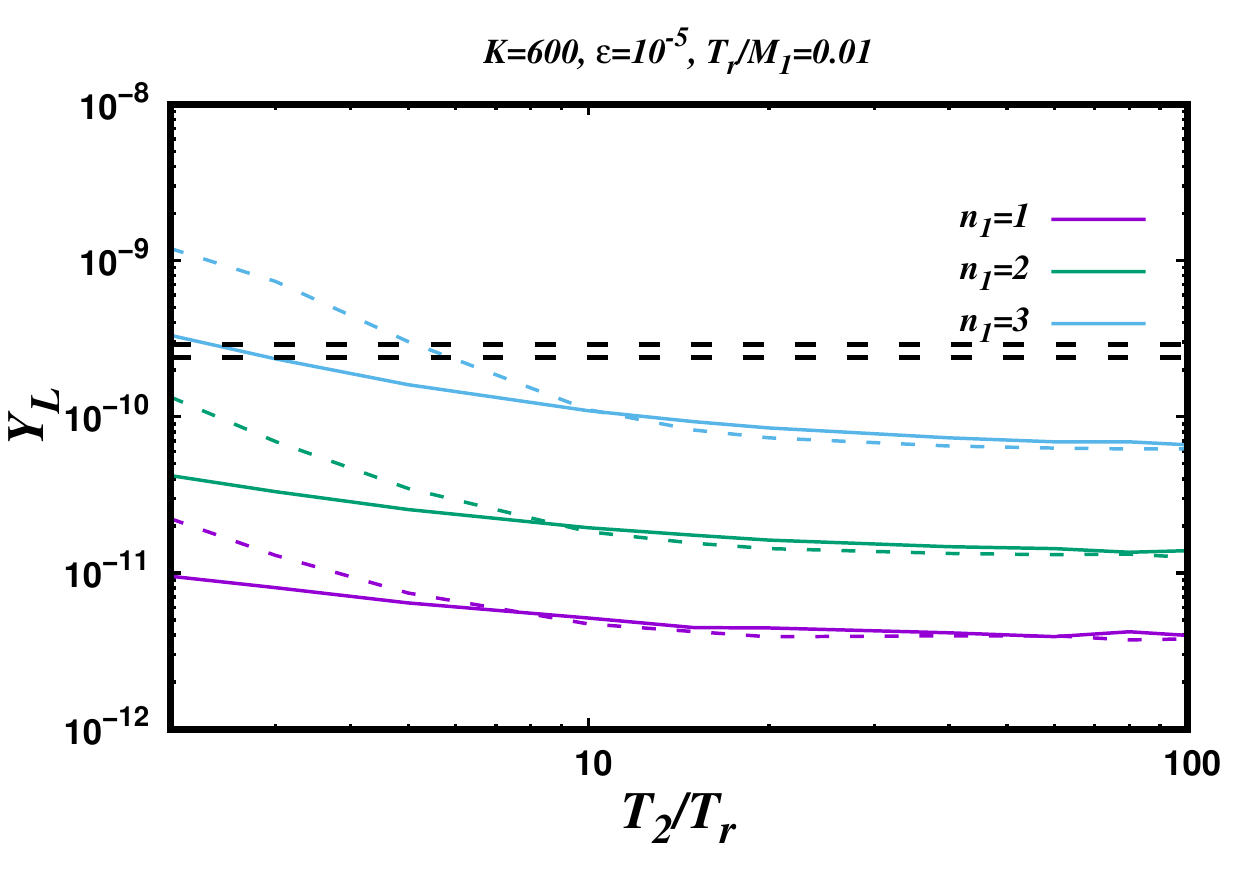}
        \includegraphics[width=0.49\textwidth]{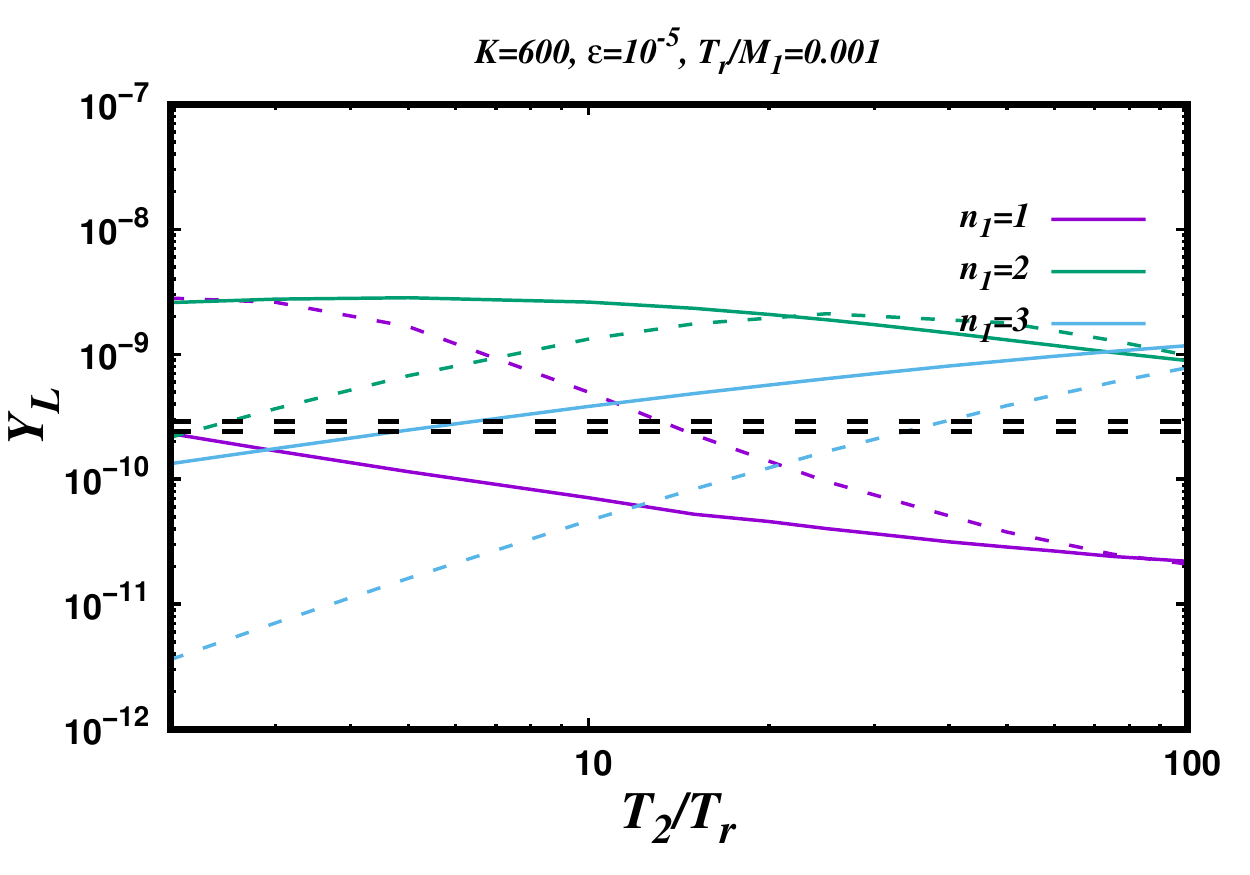}
        \caption{\it Same as Fig.~\ref{fig:2} with zero
        RHN initial abundance. The double black line(s) describes the baryogenesis threshold.}
        \label{fig:4}
    \end{center}
\end{figure}

Moreover, this behavior becomes even more prominent increasing $n_2-n_1$, 
as depicted in the right panel of Fig.~\ref{fig:3}.  
Clearly, a faster expansion with respect to the $n_2-n_1=1$ case tends to reduce 
the lepton asymmetry starting from a situation where $Y_{N_1}^{in}=0$. 
Again, it is useful to study $Y_L$ as a function of $T_2/T_r$. 
An analysis similar to that of the previous paragraph leads to the plots in Fig.~\ref{fig:4}, 
related to the case of $T_r/M_1=10^{-2}$ (left panel) and $T_r/M_1=10^{-3}$ (right panel). 
Parameters kept fixed are exactly the same as those in Fig.~\ref{fig:2}. 
For $T_r/M_1=10^{-2}$, the behavior of $Y_L$ is very similar to the previous case in Fig.~\ref{fig:2} (left panel) 
and also the analysis remains basically the same, although  $Y_{N_1}^{in}=0$. 
In the case $T_r/M_1=10^{-3}$,
when compared with Fig.~\ref{fig:2} (right panel), the quantitative result are quite different, 
but the qualitative bahavior of $Y_L$ with the ratio of $T_2/T_r$ is again basically the same.  
The reduction of the final amount of asymmetry, as mentioned, 
is due to the relevance of the inverse decay that induces oscillations in the washout mechanism.  
In any case, the requested amount of baryon asymmetry can still be obtained for a large range of $T_2/T_r$ values, 
at least when $T_r/M_1=10^{-3}$.

\section{Non-Standard History of Leptogenesis with three scalar fields}
\label{3scalar}

It is quite difficult to solve the BEs for a generic number $k\ge 3$ of additional scalar fields. 
In order to guess the trend of the solutions, it is worth to proceed with the $k=3$ example.
Already in this case modifications of the BEs for leptogenesis are complicated, with an increased number of free parameters. 
The correction factor in this case reads 

\begin{eqnarray}\label{3field}
    {\cal{J}}= 
    \left\{1+\left(\frac{M_1}{T_r z}\right)^{n_1}\left[1+\left(\frac{M_1}{T_r x  z}\right)^{(n_2-n_1)}\left(1+\left(\frac{M_1}{T_r y  z}\right)^{(n_3-n_2)} \right)\right]\right\}^{1/2}\,,
\end{eqnarray}
where $y=T_3/T_r$. 
Therefore, in the presence of three scalar field, six new parameters 
($n_i,~i=1,...,3$, $T_r$, $x$ and $y$) are necessary to introduce the modified Hubble rate. 
As described in Section \ref{Dbrane}, 
in our setting of course $T_3> T_2 > T_1=T_r$, 
with successive ordered domination from $\phi_3$ to radiation. 
Again, BEs for leptogenesis are solved for the two choices of the $N_1$ initial abundance already used in the two-scalar-field scenario.

\subsection{Case $Y_{N_1}^{in}=Y_{N_1}^{EQ}$}

\begin{figure}
    \begin{center}
        \includegraphics[width=0.45\textwidth]{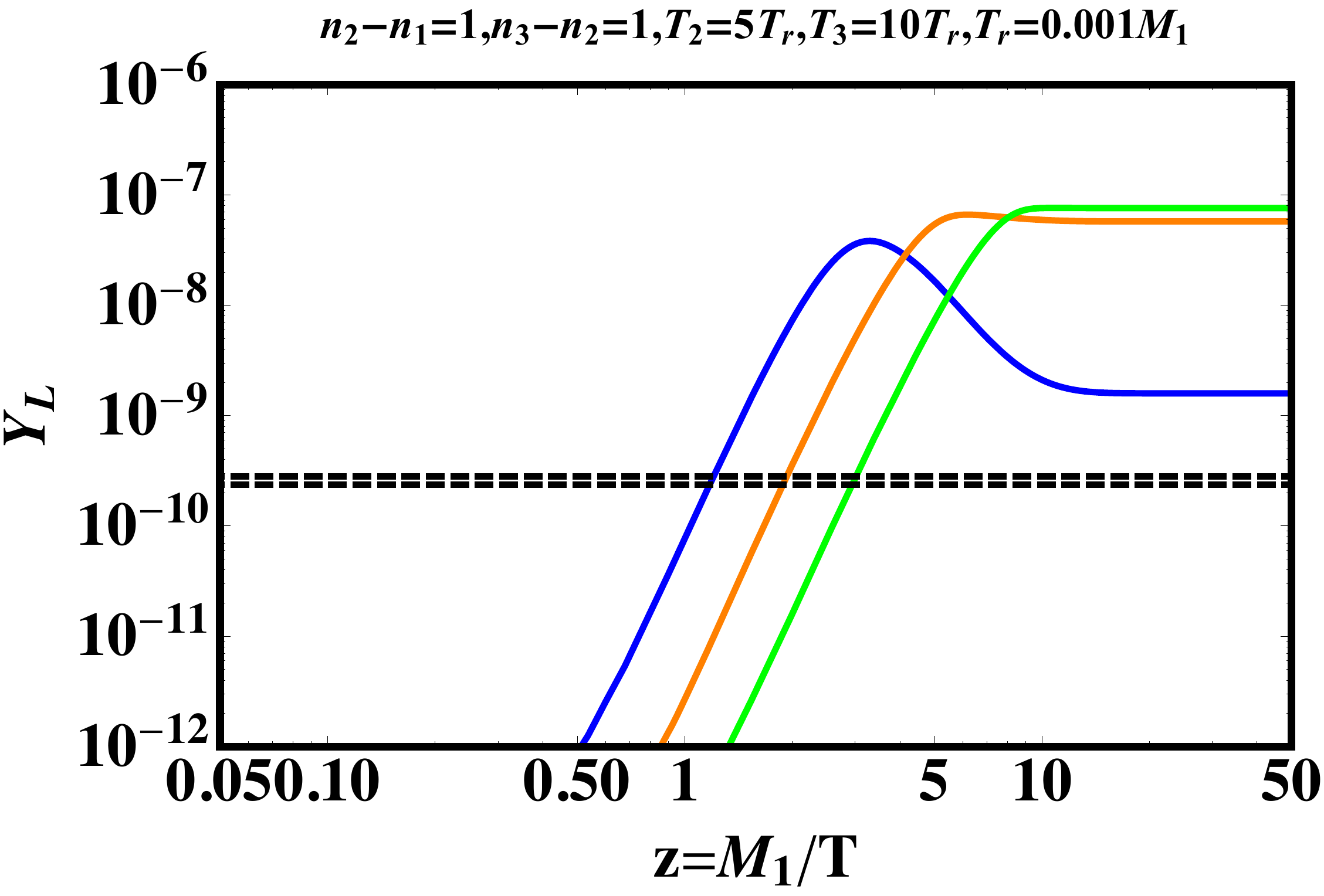}
        \includegraphics[width=0.45\textwidth]{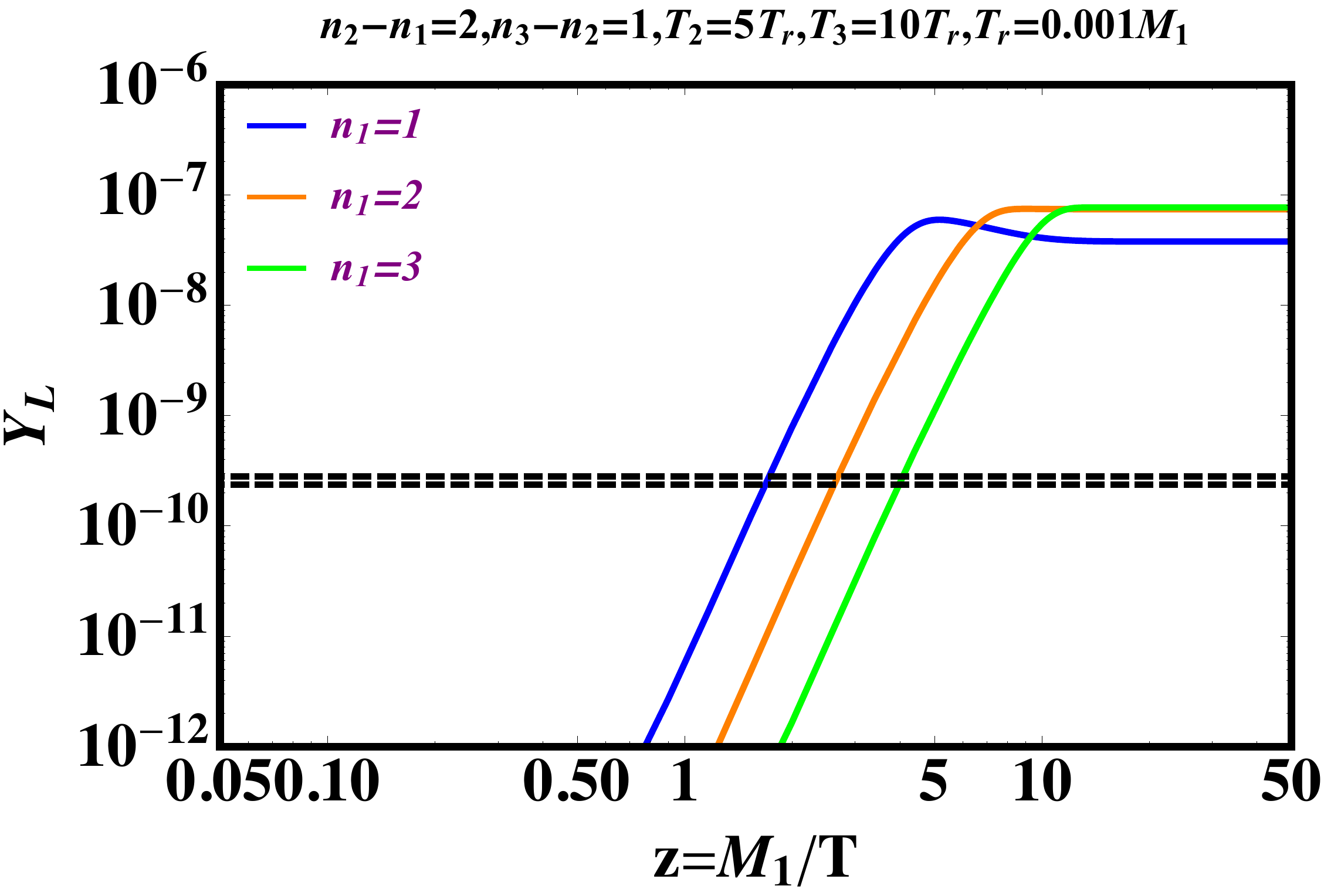}
        \caption{\it Effect of three scalar fields on $Y_L$ versus $z$ plots for initial equilibrium RHN abundance and different $n_1$ values with $n_2-n_1=1$ (left panel) and $n_2-n_1=2$ (right panel) with $T_3=10T_r$ and $n_3-n_2=1$. The double black line(s) describes the baryogenesis threshold.}
        \label{fig:5}
    \end{center}
\end{figure}

The $Y_L$ abundance is reported in Fig.~\ref{fig:5} for  initial conditions $Y_{N_1}^{in}=Y_{N_1}^{EQ}$.
The three curves in the left panel correspond to the different choices of $n_1=1,~2,~3$ and $T_3=10T_r$, $n_3-n_2=1$. 
The other parameters $\epsilon$, $M_1$, $K$, $T_2$ and $T_r$ are kept fixed at the same values  
considered in the plot of Fig.~\ref{fig:1}.  
Again, a comparison between the case with two scalar field shows that the washout effect is further reduced 
in the case where it is non-negligible, {\it i.e.} $n_1=1$. The behavior is even more significant when $n_2-n_1=2$, 
where the washout is already reduced in the presence of two scalar fields. 
The dependence on the new parameters $T_3$ and $n_3-n_2$ is also worth to be deepened, since it can change the behavior of the solutions. 
To this aim, we take 
$T_2=5 T_r$ and $T_r=10^{-3}M_1$, and solve the BEs for the four combinations of $n_2-n_1$ and $n_3-n_2$ equal to 1 or 2, 
varying also $T_3$ to be $10 T_r$, $50 T_r$ and $100 T_r$.  

\begin{figure}
    \begin{center}
        \includegraphics[width=0.45\textwidth]{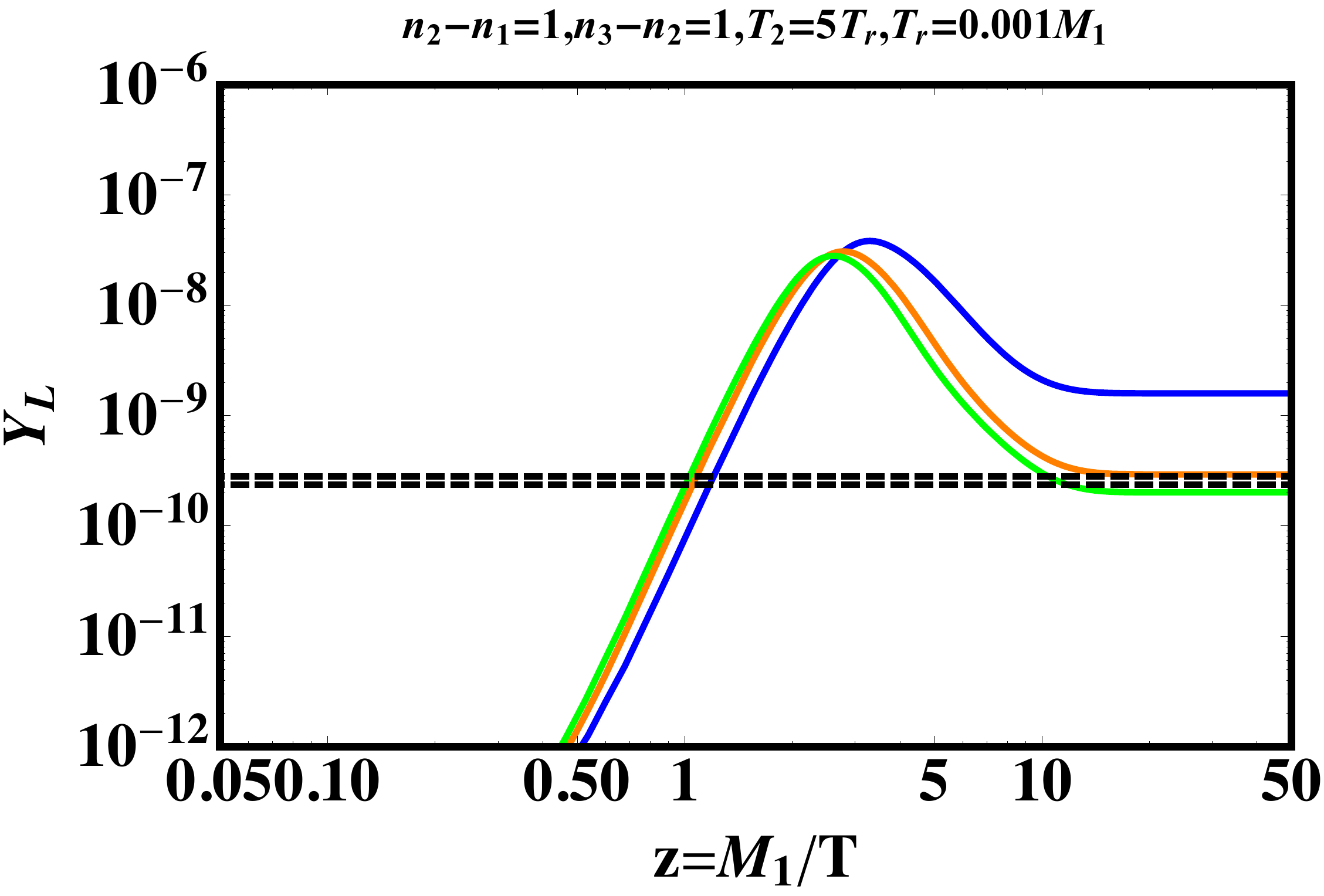}
        \includegraphics[width=0.45\textwidth]{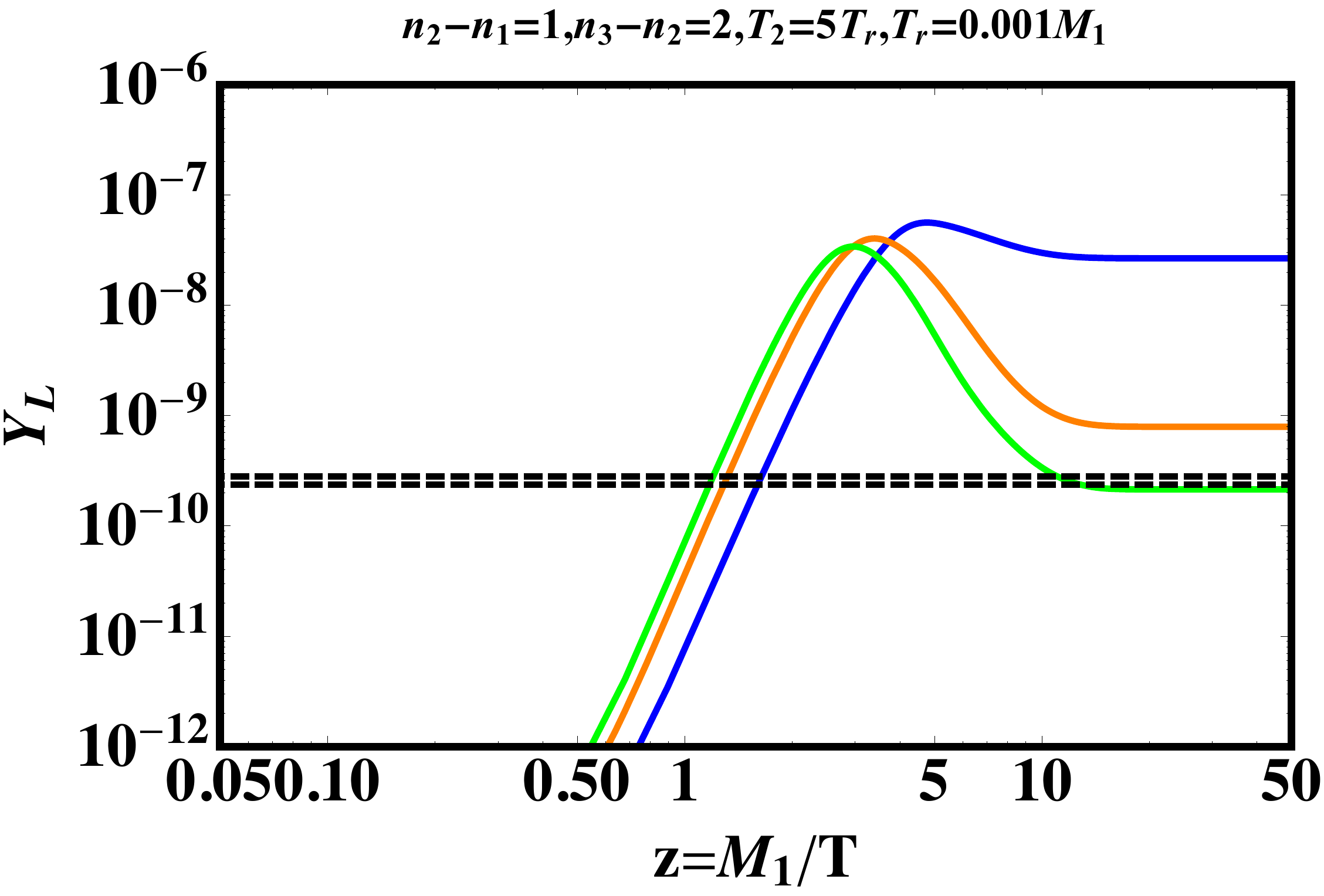}\\
        \includegraphics[width=0.45\textwidth]{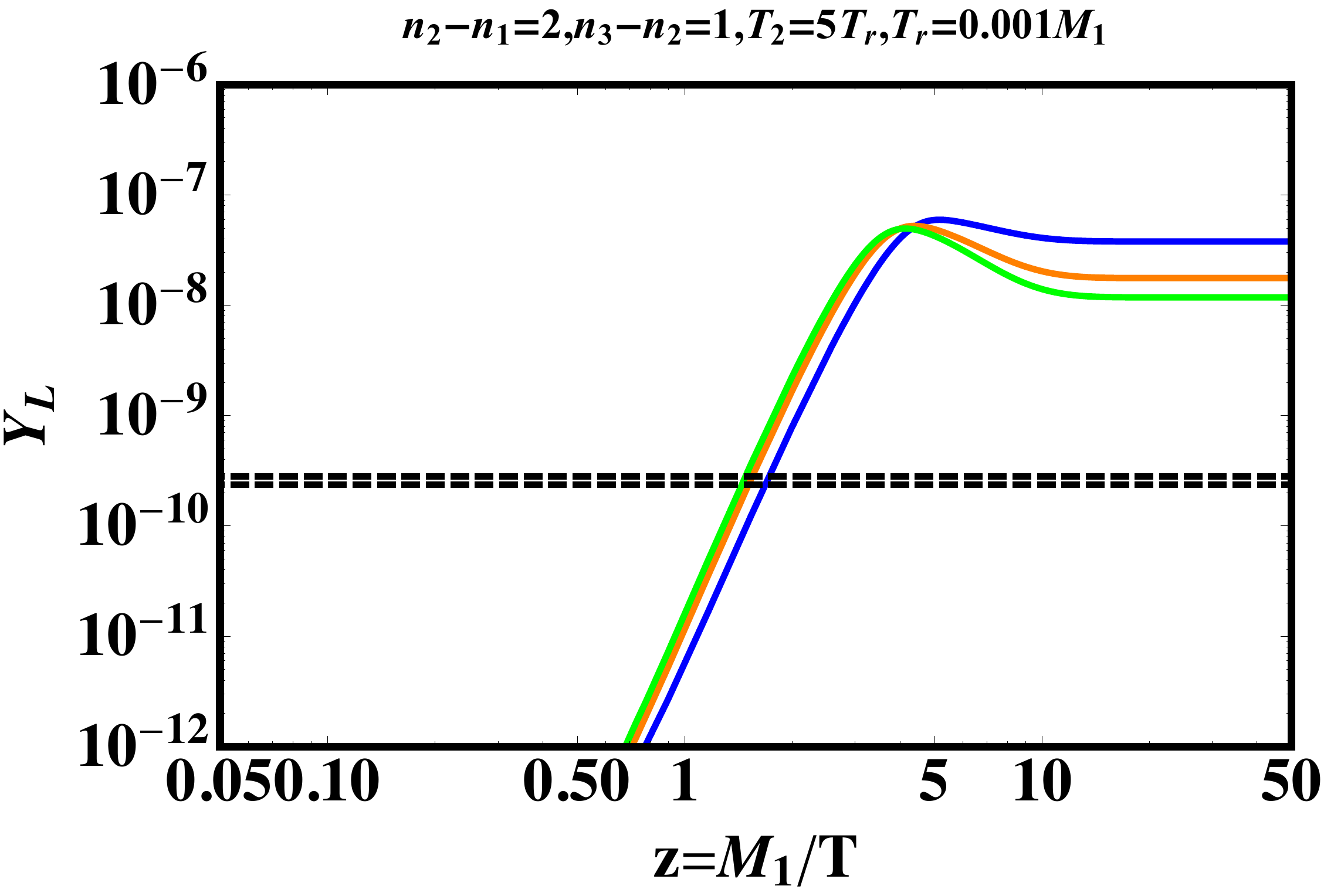}
        \includegraphics[width=0.45\textwidth]{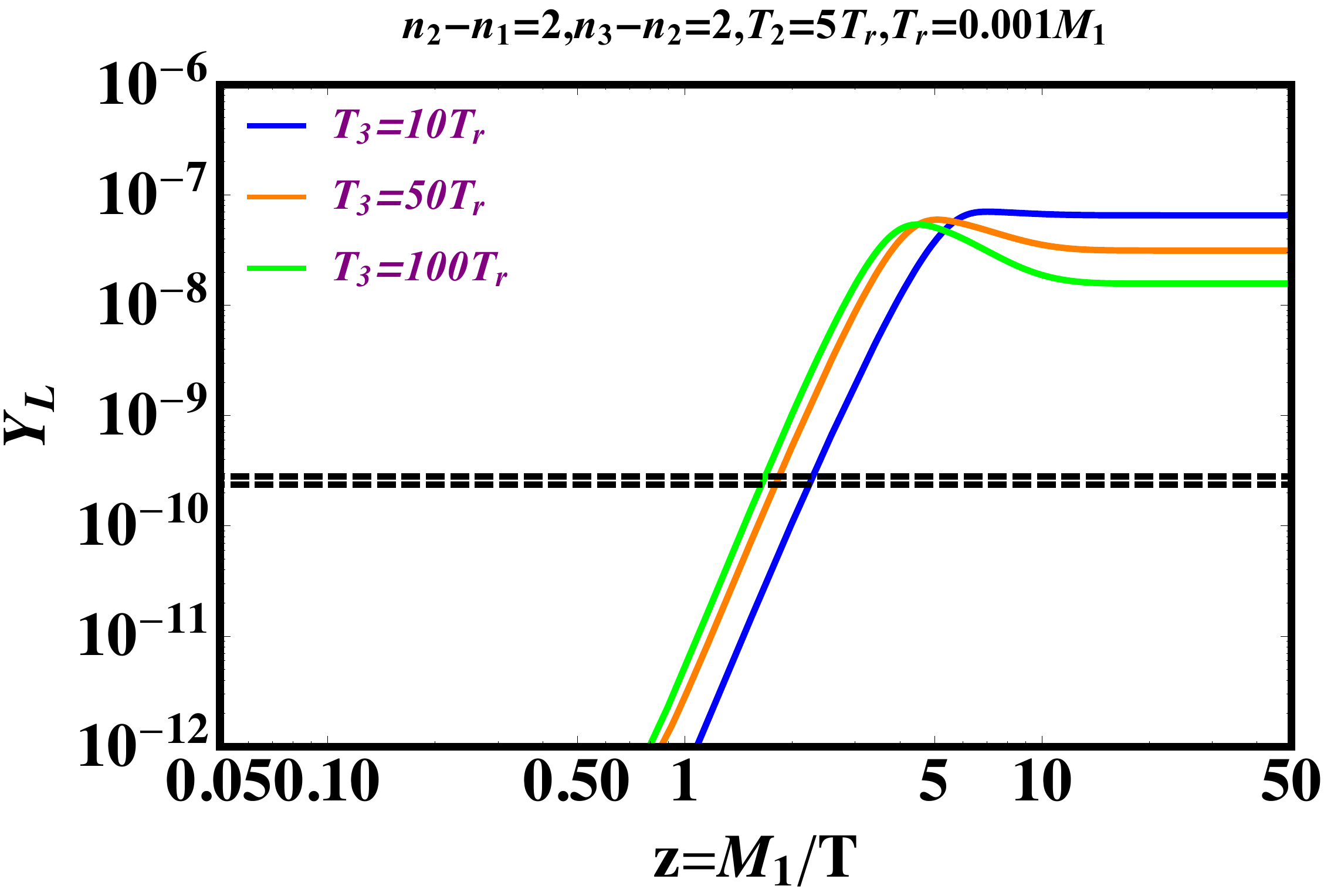}
        \caption{\it $Y_L$ versus $z$ plots for $Y_{N_1}^{in}=Y_{N_1}^{EQ}$ using various $T_3/T_r$ values ($10$, $50$ and $100$) and different $n_3-n_2$ and $n_2-n_1$ combinations of values $1$ and $2$. The double black line(s) describe the baryogenesis threshold.}
        \label{fig:6}
    \end{center}
\end{figure}

All the resulting $Y_L$ abundancies are plotted in Fig.~\ref{fig:6}. 
In the upper-left panel, it can be observed that, increasing $T_3$, the washout effect becomes prominent and the lepton asymmetry $Y_L$ decreases. 
This is obvious, because an increasing of $T_3$ makes the  third scalar insignificant. 
Indeed, for $T_3=100 T_r$, the dominance of the third scalar terminates at $T_3=M_1$, well before the decay of $N_1$,
without influencing leptogenesis.  On the contrary, for $T_3=10 T_r$, it remains effective until $T_3=0.1 M_1$, the time of leptogenesis.
A similar behavior can be observed in the other three panels, with different choices of $n_3-n_2$ and $n_2-n_1$. 
The third scalar field affects $Y_L$ when $n_3-n_2$ increases, causing lesser washout, 
as evident from the comparison of the $n_3-n_2=2$ versus $n_3-n_2=1$ cases. 
It is also clear that the influence of the third scalar depends upon the influence of the second scalar. 
In the lower panels, it can be observed that increasing $n_2-n_1$ the washout effect is sensibly reduced, 
(almost) independently on the presence of the third scalar.  
In conclusion, one may say that an increasing of both $n_3-n_2$ and $n_2-n_1$ 
results into a net decreasing of the washout effect.

\subsection{Case $Y_{N_1}^{in}=0$}

\begin{figure}
    \begin{center}
        \includegraphics[width=0.45\textwidth]{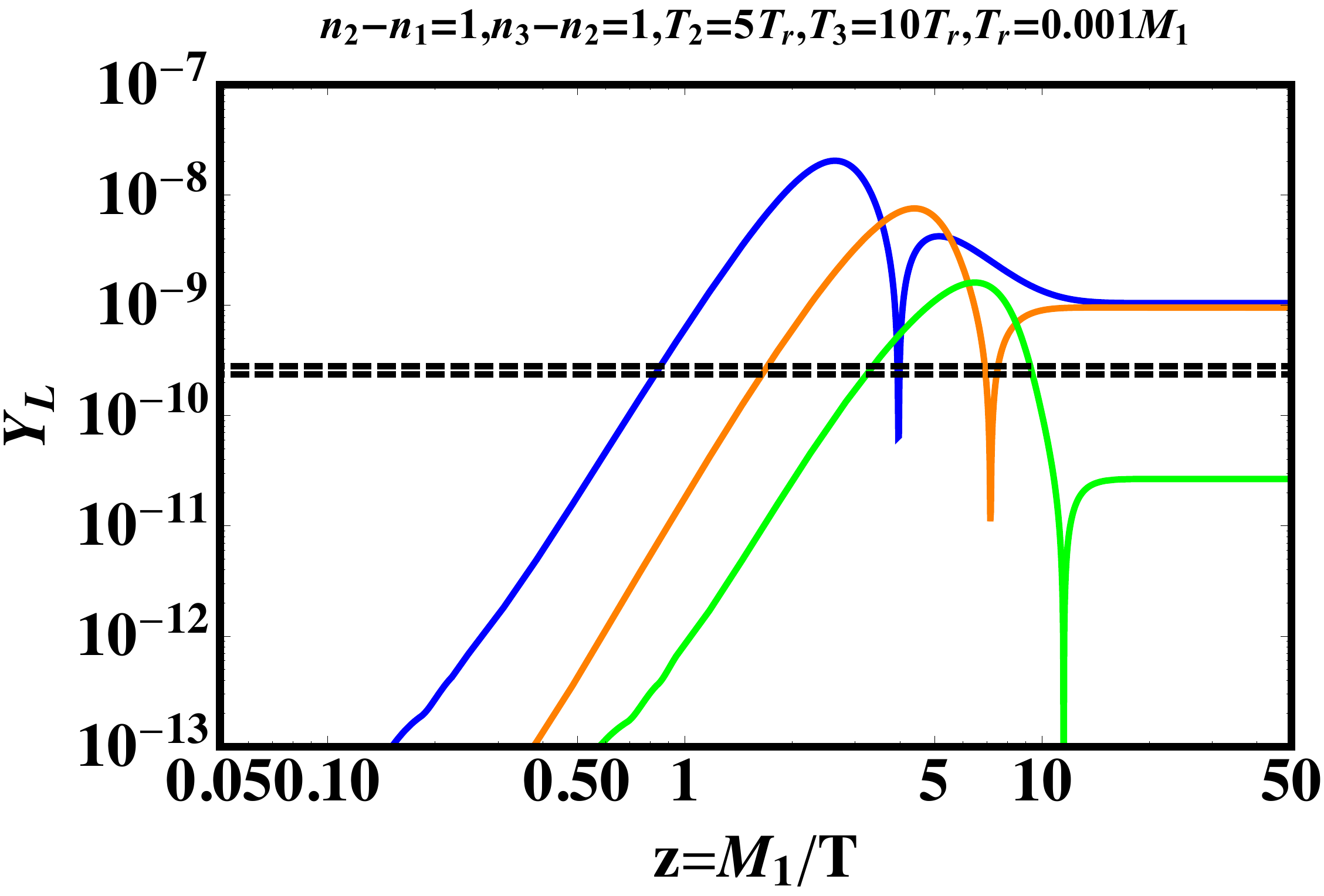}
        \includegraphics[width=0.45\textwidth]{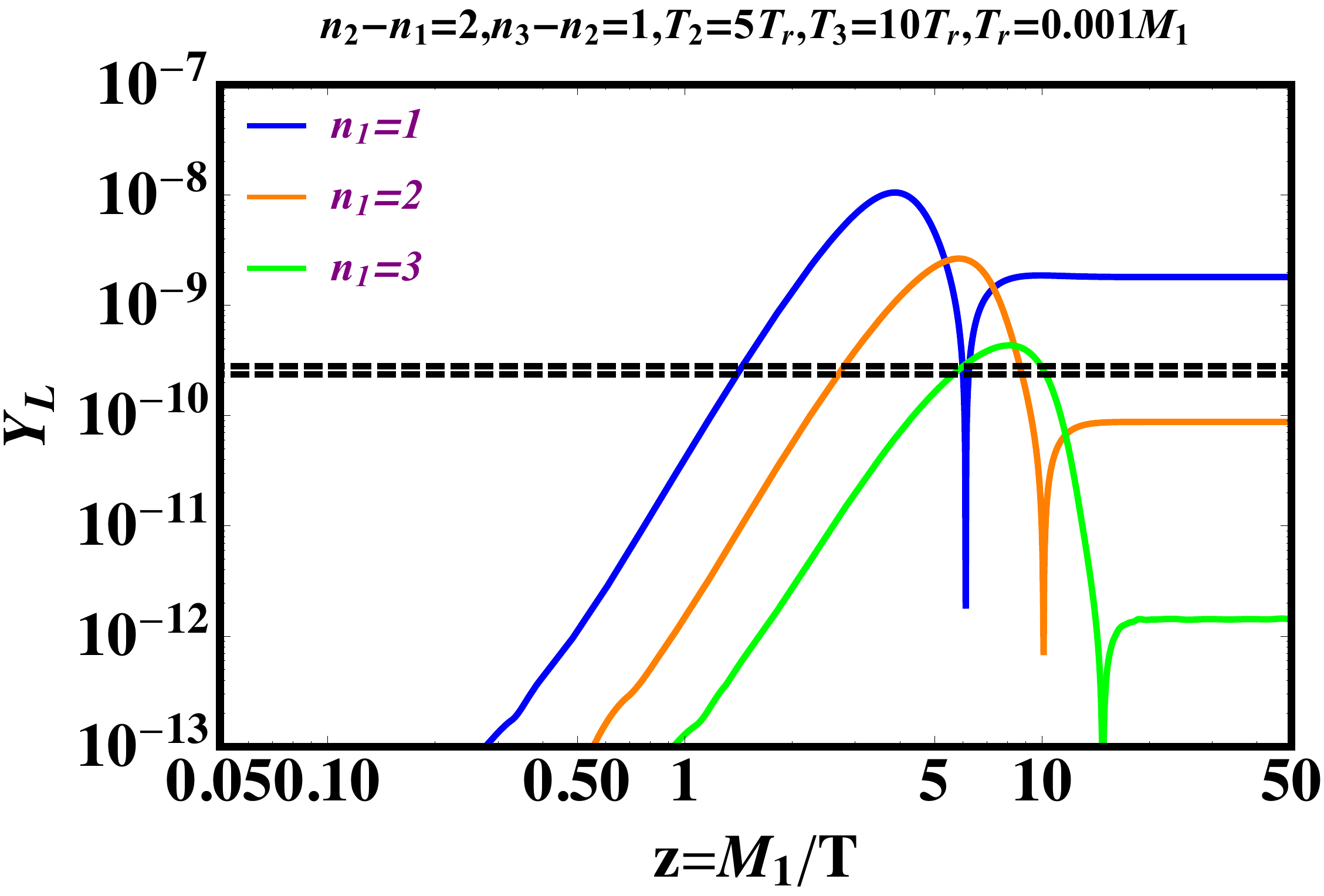}
        \caption{\it Same as Fig.~\ref{fig:5} for zero initial RHN abundance. The double black line(s) describe the baryogenesis threshold.}
        \label{fig:7}
    \end{center}
\end{figure}

Similarly to the previous discussion of leptogenesis in the presence of two additional scalar field, 
in this paragraph are analyzed the possible effects due to a third scalar field in the case of vanishing $N_1$ initial abundance. 
As before, in Fig.~\ref{fig:7} are plotted the values of 
$Y_L$ against $z$ for different $n_1$ and $n_2-n_1=1,2$.
The remaining parameters are the same as those used in 
Fig.~\ref{fig:5}.  
There is a conspicuous amount of washout for the three reported values $n_1=1, 2, 3$ 
(left panel) at the initial stage, while in the second decay the washout is limited to the $n_1=1$ case. 

\begin{figure}
    \begin{center}
        \includegraphics[width=0.45\textwidth]{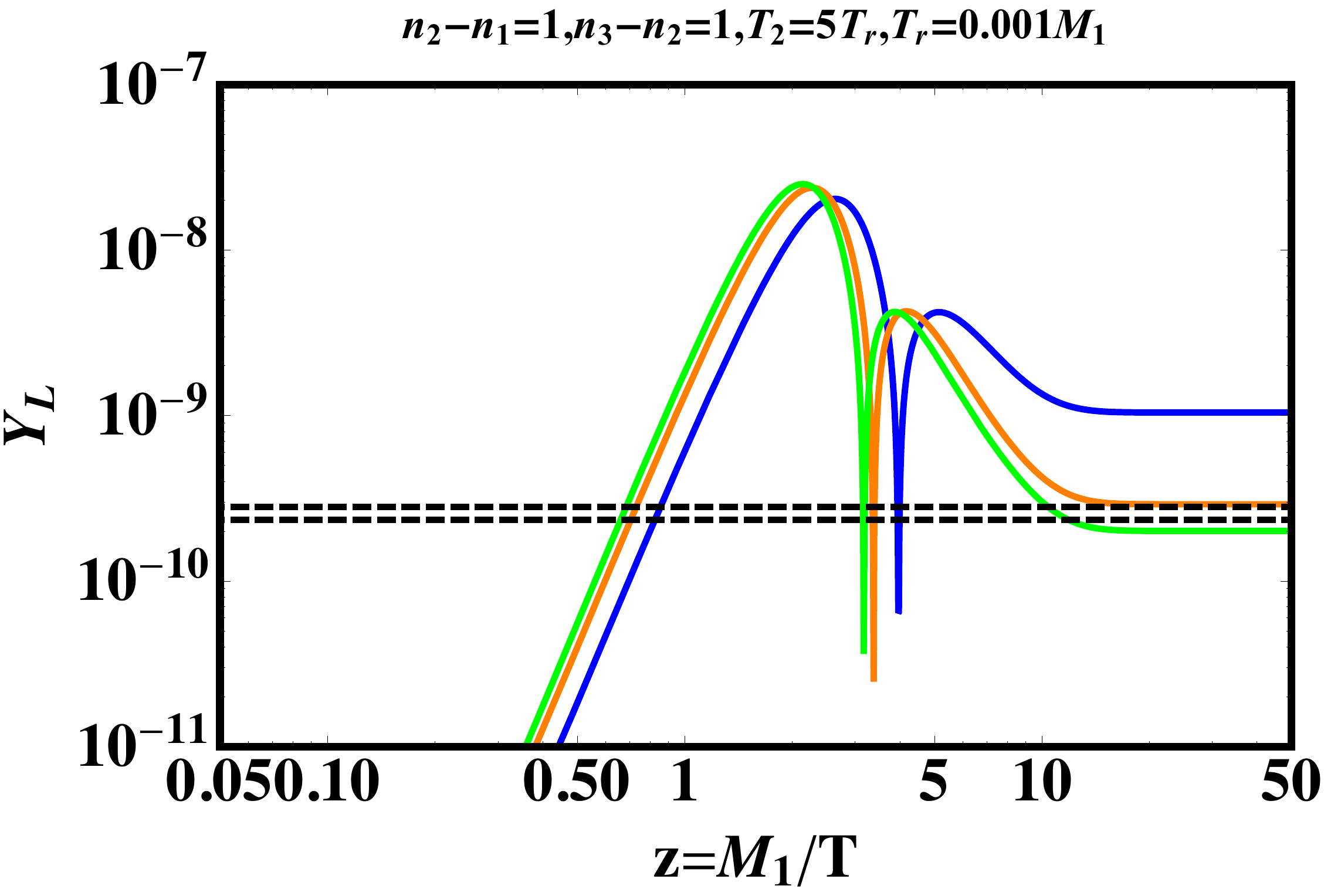}
        \includegraphics[width=0.45\textwidth]{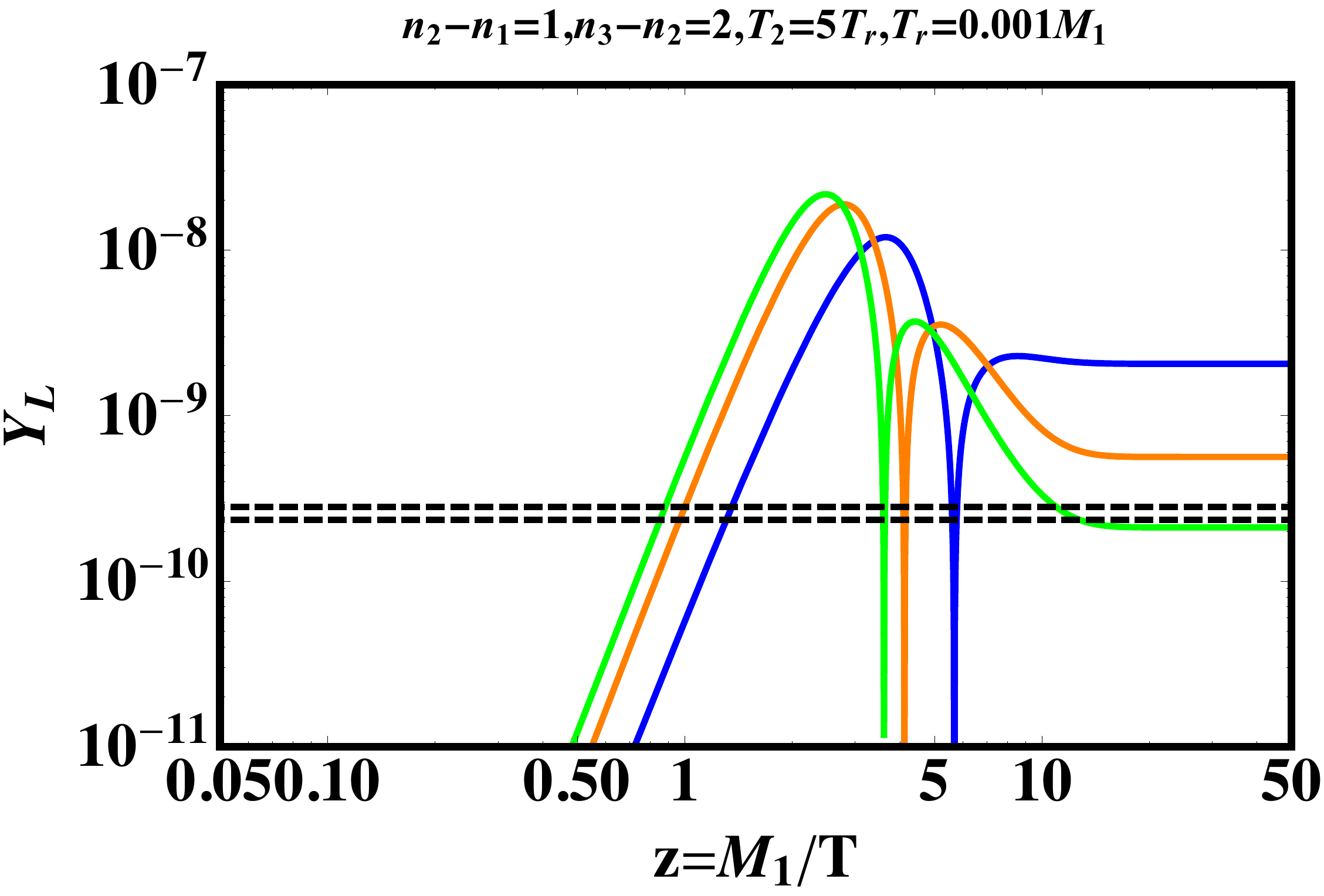}\\
        \includegraphics[width=0.45\textwidth]{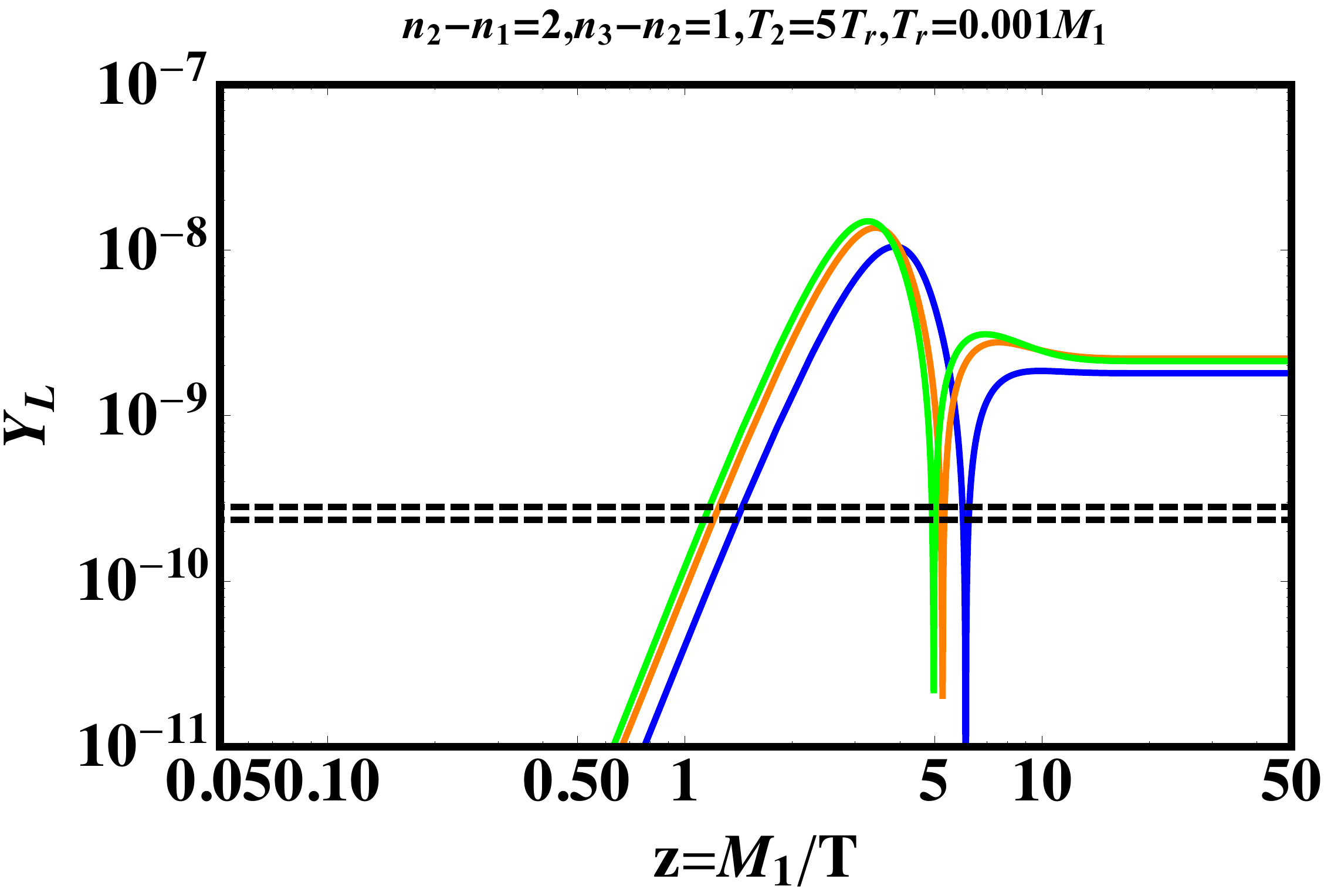}
        \includegraphics[width=0.45\textwidth]{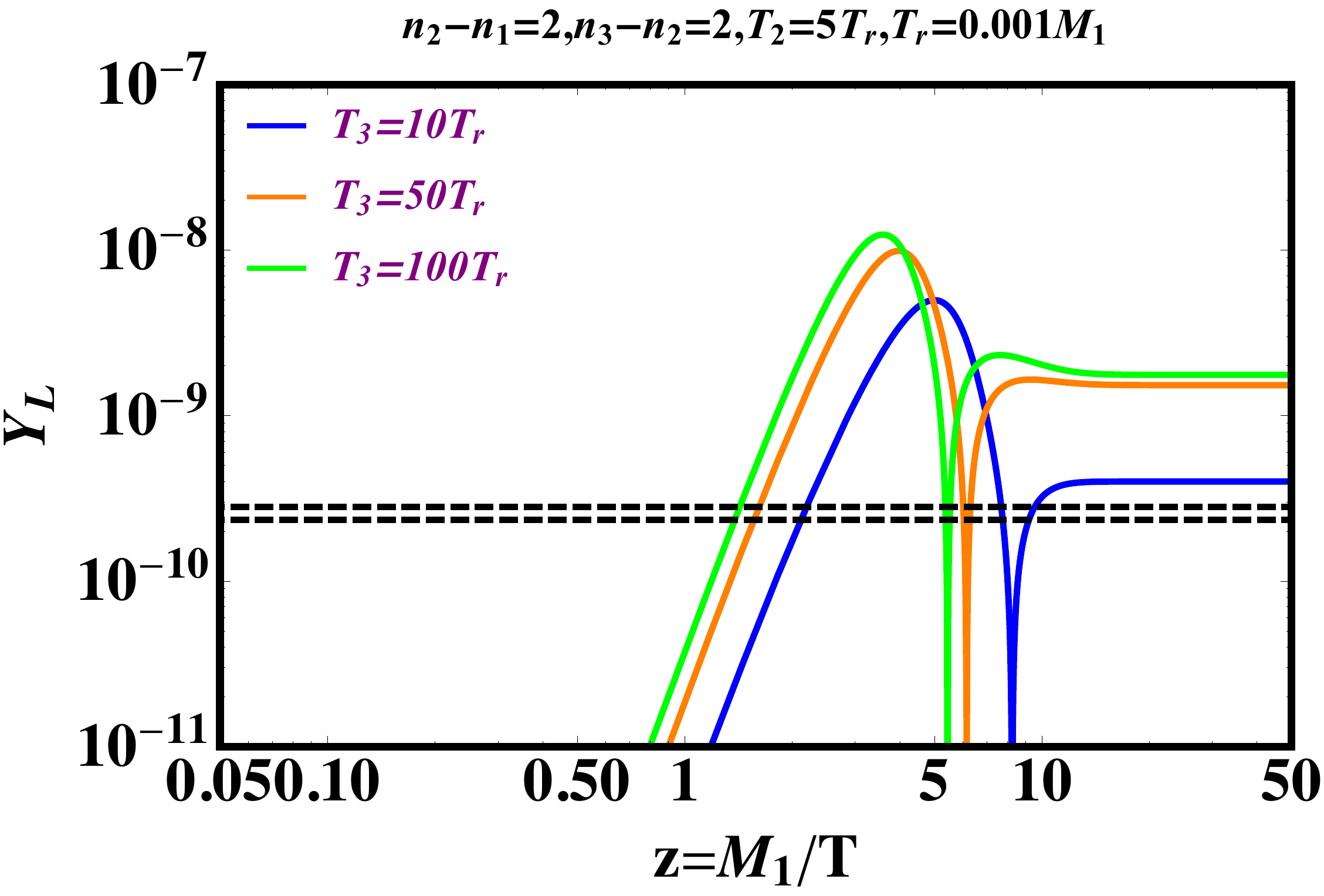}
            \caption{\it Same as Fig.~\ref{fig:6} plotted for  $Y_{N_1}^{in}=0$ with identical color scheme for different choices of $T_3$. The double black line(s) describe the baryogenesis threshold.}
        \label{fig:8}
    \end{center}
\end{figure}

Also the $Y_L$ value tends to decrease by increasing $n_1$, 
corresponding to the effect of less RHN production from inverse decay due to the faster expansion of the universe, 
as already observed in the analogous  scenario with two scalar fields. In the right panel, for $n_2-n_1=2$, 
a similar behavior is slightly milded, with the $n_1=1$ case entering a regime of weak final washout. 
It should also be noticed that, however, 
the value of $Y_L$ is reduced as the net RHN production falls down due to a weaker inverse decay in a faster expansion. 
Again, it is useful also to extend the analysis related to  Fig.~\ref{fig:6} to the case of three scalar fields 
for a variable $T_3/T_r$ and an initial vanishing $N_1$ abundance. 
An inspection of the four panels of Fig.~\ref{fig:8} 
clearly demonstrates that increasing $n_2-n_1$ or $n_3-n_2$ results into an overall dilution of the washout of asymmetry. 
In spite of it, the transition from strong to weak washout does not ensure an enhancement of the $Y_L$ value, 
governed also by the production of RHNs from the inverse decay. 
This fact can be easily extracted, for instance, by the $T_3=10 T_r$ plots (in blue) of Fig.~\ref{fig:8}. 
Initially, the lepton asymmetry enters a weak washout by changing $n_3-n_2$ 
from $1$ (upper left panel) to $2$ (upper right panel), and $Y_L$ increases. 
However, with a faster expansion ($n_2-n_1=2$, lower left panel) and even with a larger $n_3-n_2 =2$ (lower right panel), 
$Y_L$ reduces due to a compromised RHNs production. 
Things are quite different for the $T_3=50 T_r$ and $T_3=100 T_r$ plots (yellow and green curves, respectivey). 
The produced lepton asymmetry gradually enters a weak washout regime where it is enhanced (upper right and lower left panels) 
and finally saturates as the washout effect becomes negligible (lower right panel).
A behavior different from the one reported in Fig.~\ref{fig:6} and  related, as said, 
to the combined effects of RHNs production and the washout. As explained before, 
large values of $T_3/T_r$ soften the influence of the third scalar field enhancing the washout, 
being the system farer from an out-of-equilibrium phase (upper panels). 
These features disappear when the second scalar effect become stronger (lower panels). 
Finally, it should be stressed that the modifications to the BEs are basically linked to 
modifications of the effective Hubble rate, like that in eq. \eqref{Hnew} for the two-field case.  
It is thus natural to infer that our results are applicable to many other types 
of unflavored leptogenesis scenarios and are almost model-independent.

\section{Discussion and Conclusions}

It is notoriously difficult to probe the dynamics of the early universe in the epoch between cosmic inflation and the onset of BBN, 
whose predictions of primordial abundances of light elements are in very good agreement with measurements, 
and represent one of the biggest successes of modern cosmology.  
The post-inflationary evolution is thus highly unconstrained, having only to be compatible with BBN. 
In particular, all the cosmic relics that contribute to define the $\Lambda$CDM cosmological model, 
like dark matter, dark energy, baryon abundances, radiation composition and so on, crucially depend on the history around that time. 
Hence, the expansion rate can be drastically different compared to the the standard cosmology, 
in models where additional ingredients from fundamental (quantum or modified) gravity theories are present. 
For instance, four dimensional (super)string models equipped with D-branes 
typically contain additional scalar species related to the positions of the branes in the transverse internal directions, 
that in a non-equilibrium configurations could dominate the expansion rate before the radiation-dominated phase. 
It is plausible that these scalar are active during various processes in the universe 
such as post-reheating, baryon asymmetry, leptogenesis, dark matter freeze-out or freeze in, etc., 
and thus can modify significantly the thermal evolution of the universe.
In this paper, we have addressed the effects of the presence of multiple additional (sterile) scalar fields 
with a faster-than-radiation dilution law in the post-reheating epoch that, 
if active at the scale of thermal leptogenesis ($T\sim 10^{12}$ GeV), 
can cause significant changes on the baryon asymmetry (via leptogenesis) of the universe.  
In what follows, we briefly summarize our findings from the study of modified leptogenesis. 
The Boltzmann equations (BEs) describe the dynamics of the decay of the lightest RHN $N_1$, 
together with the evolution of the abundance of Lepton asymmetry $Y_L$. 
In the presence of the $k$ additional scalar fields defined in Section \ref{Dbrane}, 
the standard BEs are modified basically by the introduction of an ``effective'' Hubble rate, 
$H_{new}(T) = H(T) \, \mathcal{J}(T)$, where $\mathcal{J}$ 
is defined in eq. \eqref{eq:complmodfact}. It depends on the exponents $n_i$, on the ``separating temperatures'' $T_i$ 
and on the effective degrees of freedom active at the corresponding epochs.  
Another important ingredient is represented by the initial conditions. 
We have considered the two cases of $Y_{N_1}^{in}=Y_{N_1}^{EQ}$  (conditions ``A''), 
where the initial abundance of RHN $N_1$ coincides with the abundance at equilibrium, 
and $Y_{N_1}^{in}=0$ (conditions ``B''), with vanishing initial abundance of $N_1$. 
The initial asymmetry $Y_L^{in}$ is always taken to be vanishing. 
The main general results that can be extracted by numerical solutions of the BEs are the following:

\begin{itemize}
    \item Typically, with the increasing of $n_i$, $Y_L$ increases while the washout decreases (See Fig. 1, 3, 5, 7). 
          This is due to the fact that the faster the expansion is, the higher is the departure from thermal equilibrium.  
          As a consequence, asymmetry is feeded while washout is disfavored because there is less inverse decay.
    \item The relevance of the $\phi_{i+1}$ with respect to $\phi_i$ depends upon the difference $n_{i+1}-n_i$.  
          It clearly grows if $n_{i+1}-n_i$ increases, but $n_i$ must not be too high, otherwise the dominance of the $\phi_{i+1}$ 
          enters too early, in an epoch where the RHN $N_1$ has not been produced in a sufficient quantity (See Fig. 2, 4, 6, 8). 
          In other words, if $\phi_i$ already absorbs the whole washout, the $\phi_{i+1}$ action ceases to be significant.
    \item In the evaluation of the $\phi_i$ contribution to leptogenesis, $T_i$ is of course a fundamental parameter, 
          since the field $\phi_i$ is active only if $T_i<M_1$. Moreover, if the ratio $T_{i+1}/T_i$ decreases, 
          the $\phi_{i+1}$-domination epoch is longer and $Y_L$ becomes bigger. 
          All the temperatures have to be related also to $M_1$ and to $T_1=T_r$.
    \item With the $Y_{N_1}^{in}=Y_{N_1}^{EQ}$ initial conditions, 
          the production of asymmetry $Y_L$ is typically monotonic and after a washout the value of $Y_L$ saturates at a certain value. 
          To evaluate if leptogenesis is so efficient to generate the requested amount of baryon asymmetry, 
          one has to analyze the balance between the values of the dilution exponents $n_i$ and the ratios of the temperatures $T_i$ 
          to the radiation temperature $T_r$ (Fig. 1, 2, 5, 6). 
    \item With the  $Y_{N_1}^{in}=0$ initial conditions, there is an oscillation due to the strong initial washout, 
          since the inverse decay of the produced RHN $N_1$ is large at the beginning and starts with a vanishing initial abundance.  
          The saturation of $Y_L$ at a certain value is thus slower and the amount of asymmetry $Y_L$ can be small.
          As in the previous case, in order to understand if leptogenesis can generate baryogenesis, 
          one has to evaluate the dipendence of $Y_L$ upon the $n_i$ and the temperatures $T_i$ (Fig. 3, 4, 7, 8). 
    \item It is quite clear, however, that in general more scalar fields contribute to increase the value af the asymmetry $Y_L$ 
          and in the worst case become uneffective for the reasons mentioned above.  However with vanishing initial $N_1$ abundance, 
          due to a weaker inverse decay in a faster expansion, there is a reduction of the saturated $Y_N$ value.
    \item We have studied in details the case with two scalar fields, 
          where it is indeed possible to satisfy Baryon asymmetry in the universe within the range $0.001\le T_r/M_1 \le 0.01$ 
          for thermal leptogenesis with a chosen set of parameters $M_1=10^{11}$ GeV, $\epsilon=10^{-5}$ 
          and $K=600$ (see Section \ref{leptogen}) for a large interval of $T_2/T_r$ values (Fig. 2, 4).
    \item In the case of three scalar fields, also studied in details, 
         it is important to analyze the behavior of the system with initial conditions $Y_{N_1}^{in}=0$  
         in comparison with  the  $Y_{N_1}^{in}=Y_{N_1}^{EQ}$ initial conditions. 
         Again, as in the presence of two-scalar fields, a decreasing of the washout accompanied by a decreasing of 
         the $Y_L$ values can be observed. 
         Moreover, one finds that leptogenesis can generate baryogenesis in the interval $10\le T_3/T_r \le 100$ 
         for different $n_2-n_1$ and $n_3-n_2$ (see Section \ref{3scalar} and Fig. 6, 8, for details). 
\end{itemize}    

It is important to notice that since the modifications in leptogenesis 
are obtained by changing the Hubble parameter into the effective one, we expect that our findings be
applicable to any other thermal leptogenesis models,  
independently of the choice of the seesaw mechanism.  
Therefore, one can actually obtain a different regime of consistent parameter space in model dependent studies 
due to the influence of these scalar fields. 

The scale of leptogenesis in the case of a typical type I seesaw model discussed in the present work, 
is very high and is out of reach for the ongoing experimental facilities because the RHN mass be above $10^9$ GeV \cite{19}. 
In general, one may only get indirect signals for leptogenesis 
via observations of neutrino-less double beta decay \cite{25}, 
via CP violation in neutrino oscillation \cite{26}, 
by the structure of the mixing matrix \cite{27}, 
or from constraints relying on Higgs vacuum meta-stability in the early universe \cite{28,29}, 
that pin down very tightly the parameter space of heavy neutrino physics \cite{30}. 

Several mechanisms with a much lower leptogenesis scale exist where the RHN masses arise due to new physics around the TeV scale 
if two RHNs are nearly degenerate in mass, known as \emph{resonant leptogenesis} \cite{31}
or via oscillations of GeV-scale right-handed neutrinos \cite{32}
or via Higgs decay \cite{33} 
as well as dark matter assisted scenario with $1$ to $3$-body decays \cite{34}
which allow these models to be probed at the ongoing experimental facilities. 
In such models, it is plausible to obtain  successful leptogenesis 
via RHN with mass $M_1 \sim$ 10 TeV assuming an initial thermal abundance for RHNs along with an almost absence of washout. 
It is possible that the scalar fields discussed in the present work may remain active at low scale as well. 
Therefore, we expect significant deviation from the results reported in such low scale leptogenesis models 
once a different thermal expansion is invoked via scalars but remaining above the energy scale 
where the sphalerons are active to transfer the asymmetry to the baryon sector.  
Moreover, possible observations of primordial gravitational waves sourced by topological defects \cite{35}, 
colliding vacuum bubbles \cite{36} 
primordial black holes \cite{37} 
and Cosmic Microwave Background radiation (CMBR) measurements \cite{38}
should represent additional and complementary tools to probe leptogenesis at high energy scales.
However, it should be stressed that effective models, 
like the one presented here, 
are not easy to be distinguished by other mechanisms providing similar effects. 
It would thus be interesting to understand if there can be experimental probes to the proposed scenario.

\section{Acknowledgements}

ADB acknowledges financial support from DST, India, under grant number IFA20-PH250 (INSPIRE Faculty Award). 
This work is supported in part by the DyConn Grant of the University of Rome ``Tor Vergata''.

\end{document}